\documentclass[a4paper,fleqn,usenatbib]{mnras}

\usepackage{newtxtext,newtxmath}

\usepackage[T1]{fontenc}
\usepackage{ae,aecompl}

\usepackage{graphicx}	
\usepackage{amsmath}	
\usepackage{amssymb}	

\title[Early multiwavelength analysis of V745 Sco]{Early multiwavelength analysis of the recurrent nova V745 Sco}

\author[Delgado \& Hernanz]{
L. Delgado,$^{1,2}$\thanks{E-mail: delgado@ice.cat}
M. Hernanz,$^{1,2}$\thanks{Corresponding author; E-mail: hernanz@ice.csic.es}
\\
$^{1}$Institut de Ci\`encies de l'Espai (ICE, CSIC), Campus UAB C/ Can Magrans s/n, 08193 Bellaterra (Barcelona), Spain\\
$^{2}$Institut d'Estudis Espacials de Catalunya (IEEC), 08028 Barcelona, Spain
}

\date{Accepted XXX. Received YYY; in original form ZZZ}

\pubyear{2018}

\begin{document}
\label{firstpage}
\pagerange{\pageref{firstpage}--\pageref{lastpage}}
\maketitle

\begin{abstract}

In recent years, several nova explosions have been detected by Fermi/LAT at E$>$100 MeV, mainly early after the explosion and for a short period of time. The first evidence of particle acceleration in novae was found in the 2006 eruption of RS Oph, to explain the faster than expected deceleration of the blast wave. As a consequence, emission of high-energy gamma-rays mainly from neutral pion decay and inverse Compton scattering is expected. We aim to understand the early shock evolution, when acceleration of particles can take place, in nova explosions. To achieve this goal, we perform a multiwavelength study of the 2014 outburst of V745 Sco, a symbiotic recurrent nova similar to RS Oph. The analysis of early Swift/XRT observations, simultaneous to the tentative Fermi detection, is combined with Chandra and NuStar data, to get a global picture of the nova ejecta and the red giant wind evolution. Early radio and IR data are also compiled, providing information about the forward shock velocity and its magnetic field. The comparison with the plasma properties of RS Oph shows striking similarities, such as the skipping of the adiabatic phase occurring in supernova remnants, a hint of particle acceleration. The multiwavelength study of V745 Sco provides new insights into the evolution of the hot plasma in novae and its interaction with the circumstellar material, a powerful tool to understand the nature of the high-energy gamma-ray emission from symbiotic recurrent novae.

\end{abstract}




\section{Introduction}

In recent years several novae have been detected at E$>$100 MeV by the \textit{Fermi} satellite, demonstrating that particles (p and e$^{-}$) are accelerated to relativistic energies due to strong shocks in the ejecta. In most cases,  high energy emission has been observed early after the explosion and for a short period of time \citep{Abdo2010,Cheung2016}. This emission is a consequence of $\pi^{0}$  decay (hadronic process) and/or Inverse Compton scattering (leptonic process). Particle acceleration in novae was inferred for the first time in the 2006 outburst of RS Ophiuchi \citep{Tatischeff2007}. The authors showed that the blast wave decelerated faster than expected as a consequence of acceleration of particles in the shock. 

In some novae hard X-ray emission is detected the first days after the outburst \citep{Sokoloski2006}. The shock wave running into the circumstellar material 
heats the material, which emits as an optically thin plasma in collisional ionisation equilibrium. The regions that emit X-rays reach extremely high temperatures ($\sim$10$^{8}$ K) at which it is not expected to detect emission in IR lines such as Pa $\beta$ or O I lines. According to \citet{Das2006}, the near-IR emission detected in some symbiotic recurrent novae is originated in the contact discontinuity between the ejected material (hot compressed region) and the RG wind (cool uncompressed region), a cooler and denser region propagating with the same speed as the shock front \citep{Lamers1999}. The radio emission is dominated by thermal bremsstrahlung (free-free emission) in most novae. However, in some cases synchrotron non-thermal emission, related with a significant population of relativistic electrons accelerated in the shocks, is also observed  \citep{BodeEvans2008, Chomiuk2014}.  Therefore, IR and radio observations give insights into particle acceleration in novae.

The hard X-ray emission and the radio emission observed in the classical novae (CNe) detected by Fermi, were not simultaneous with the Fermi detection (except for the recent cases mentioned below). It was even necessary to wait for weeks or months after the outburst to obtain the first detection in hard X-rays. Then, it is not possible to relate these hard X-rays with the shock that produces the acceleration of particles in classical novae. \citet{Metzger2014} state that the X-ray emission produced in the shock is not observed in the classical novae detected by Fermi because it is absorbed and reprocessed to the optical range. This is due to the high column densities in the ejecta. In the case of recurrent novae (RNe) the ejecta densities are expected to be smaller than in classical novae because less mass is ejected.  So far, four classical novae, V339 Del,V5668 Sgr, V5855 Sgr and V906 Car(=ASASSN-18fv), have been observed at E$>$10 keV by \textit{NuSTAR} during their Fermi detection; two of them have been detected by  \textit{NuSTAR} : V5855 Sgr  \citep{Nelson2019} and V906 Car \citep{NelsonATel2018}, whereas for V339 Del and V5668 Sgr only upper limits have been obtained \citep{Vurm2018}.

This work focuses on the study of the symbiotic recurrent nova V745 Sco, that  was tentatively detected by \textit{Fermi} in its last outburst in 2014, with 3$\sigma$ significance level only, probably because of its long  distance (about 8 kpc). V745 Sco is a symbiotic binary system with a red giant (RG) companion. In symbiotic binary systems, the white dwarf (WD) captures matter from the stellar wind of the RG; moreover, matter is presumably also transferred via an accretion disc, as in CNe. Their orbital periods are around 100 days, longer than the typical periods of cataclysmic variables, and the corresponding orbital separations are $10^{13}-10^{14}$ cm  \citep{Anupama1999}. The short recurrence periods of RNe  (P$_{\text{rec}}<100$ yr) are explained theoretically with a highly massive WD, probably close to the Chandrasekhar limit, together with a high accretion rate ($\sim 10^{-8}-10^{-7}$ M$_{\odot}$yr$^{-1}$). For this reason, these novae are candidates for SNeIa explosions.

The aim of this work is to study the evolution of the symbiotic recurrent nova V745 Sco the first days after outburst through a multiwavelength study, with special emphasis on X-ray emission, to better understand particle acceleration, which is responsible for the production of the high-energy gamma-rays detected by \textit{Fermi}. In \S 2 we describe  the characteristics of V745 Sco and  give an overview of its 2014 outburst. In \S 3 early X-ray observations with different instruments are presented and analysed in detail, whereas in \S 4 a reanalysis  of IR observations is carried out. The properties and evolution of the plasma behind the forward shock and the RG wind are presented in \S 5 and \S 6. The comparison with RS Oph and the implications for particle acceleration are discussed in \S 7. Conclusions and a summary of the results are included in \S 8.

\section{The 2014 outburst of V745 Sco}\label{sec:V745Sco}

V745 Sco is one of the ten recurrent novae  known in our galaxy and it has been detected in eruption three times. V745 Sco is a symbiotic binary system formed by a WD and a M-type RG \citep{Duerbeck1989,Harrison1993}. 
Its first known outburst, in 1937, was discovered by \citet{Plaut1958}, on plates taken at the Leiden Observatory. The second detection was in 1989, by \citet{Liller1989}, 
as referred in the extensive review of recurrent novae by \citet{Schaefer2010}. Schaefer suggested an additional unobserved outburst around 1963 between the two first detections and proposed a recurrence period of $\sim$25 years. This period was confirmed 25 years later with a new outburst in 2014. This nova decreases in brightness very fast (t$_{3}$=7 days) and has a relatively faint peak (m$_{\text{v}}$=9 mag in the maximum). For these reasons, some outbursts might have been missed \citep{Page2015}. 

\citet{Anupama2008} classified V745 Sco as a long period recurrent nova system - like RS Oph, T CrB and V3890 Sgr - based on its estimated orbital period. There are some discrepancies about it, but it is known that if the secondary is a red giant, the orbital period should be a few hundreds of days. The optical observations in quiescence taken at Cerro Tololo between 2004 and 2008 by \citet{Schaefer2009,Schaefer2010} indicated a photometric periodicity of 255 $\pm$ 10 days. 
However,  \citet{Mroz2014} showed that there is no 255 days-periodicity or ellipsoidal effect, but instead they found semi-regular pulsations of the RG with sinusoidal variations with 136.5 days and 77.4 days period.  
Also, in the last outburst in 2014 periods of 77 and 155 days were observed with SMARTS  \citep[F. Walter priv. comm.; see also][]{Page2015}. 

The distance to V745 Sco is not clear but \citet{Sekiguchi1990} suggested that this nova must be in the Galactic Bulge due to its position close to the Galactic Center. Therefore, its distance must be about 8 kpc. Using different methods to calculate the distance through the effective surface temperature, \citet{Schaefer2010} found a distance of 7.8$\pm$1.8 kpc that is in agreement with the suggestion of \citet{Sekiguchi1990}. We will adopt this value in this work.

\begin{table}

\caption{Observations of V745 Sco}
\label{tab:observaciones_V745Sco}

\scalebox{0.83}{\begin{tabular}{ l l l l }

\hline
\hline
  Days after& Wavelength 		&Telescope  	& Reference\\
discovery	&  					&				& 					\\

\hline
\hline
  0	&		Optical	&							&  AAVSO$^{1}$	\\
1.3-15.3 	&     Optical	&2.3m Vainu Bappu   & \citet{AnupamaATel2014} \\
1.7-42	&		Optical	&			SMARTS		&  \citet{Page2015}	\\
	\hline
 1-2			& $\gamma$-rays& \textit{Fermi}& \citet{CheungATel2014} 	\\
 		&			&	(upper limits)					   & \citet{Cheung2016} 	\\ 
 \hline
 0.16-229.31 &X-rays & \textit{Swift}/XRT	   & \citet{Page2015}	\\ 
 		&			&						   &  \citet{Drake2016}	\\
10 		&X-rays			&\textit{NuStar}	   & \citet{Orio2015}	\\
16 		&X-rays			&\textit{Chandra}	   &\citet{Drake2016}\\

 \hline
 1.3-15.3	&IR		&1.2m Mount Abu	           & \citet{Banerjee2014}	\\
 \hline
 1-217 	&Radio			&GMRT		           &  \citet{Kantharia2015}	\\
2-3 		&Radio			&VLA			   & \citet{RupenATel2014}	\\

\hline
\hline
\end{tabular}}

\end{table}

The last outburst of V745 Sco was discovered on 2014 February 6.694 UT by Rob Stubbings  \citep[AAVSO Alert notice \#380\footnote{\url{ https://www.aavso.org/aavso-special-notice-380}}; see also][]{Waagen2014} with  m$_{v}\approx9$~mag. For the purpose of this work, we define this as t=t$_{0}$. There was no evidence of the nova 24 hours before, with a limiting  m$_{\text{v}}\approx$ 13.0 mag. This last eruption was detected  from $\gamma$-rays to radio.  V745 Sco is the first RS Oph-like nova detected at very high energies and the sixth nova detected with \textit{Fermi}/LAT. Detections took place 1 and 2 days after the outburst (2014 February 6 and 7) with 2$\sigma$ and 3$\sigma$ significances. Four days later, only upper limits were  obtained \citep{CheungATel2014,Cheung2016,Franckowiak2017}. 

In near IR, \citet{Banerjee2014} found broad lines with a narrow peak and expansion velocities around 4000 km s$^{-1}$ (FWZI$>$9000 km s$^{-1}$) similar to the line profiles observed in the optical by \citet{AnupamaATel2014}. 

Radio emission was reported by \citet{RupenATel2014}, who detected a rising spectrum with VLA 2 days after the optical maximum, that might be consistent with thermal emission or self-absorbed synchrotron emission, while \citet{Kantharia2015} detected synchrotron emission at 610 MHz and 235 MHz with GMRT 26 days after the eruption. The magnetic field of V745 Sco was estimated to be 0.03 G, similar to that of RS Oph (0.04 G). These observations are listed in Table \ref{tab:observaciones_V745Sco}. 

\section{X-ray observations}

During its last outburst in 2014, V745 Sco was observed with several X-ray satellites covering all the evolutionary stages of the nova. The \textit{Swift} satellite continuously monitored the source during seven months, until V745 Sco was too dim (Table \ref{tab:observaciones_V745Sco}). The first days after the outburst, it was detected with \textit{Swift}/XRT  (see \S \ref{sec:early_observations}).  On day 4, the nova entered in the SSS phase; a detailed study of this phase is reported in \citet{Page2015}. On the other hand, \textit{Swift}/BAT did not detect V745 Sco in the 15-50 keV energy range \citep[H.A. Krimm, priv. comm.; see also][]{Page2015}. V745 Sco was also detected with \textit{NuStar}, 10 days after outburst in the energy range 3-20 keV; but only upper limits could be derived in the 20-70 keV range \citep{Orio2015}. Also \textit{Chandra} detected V745 Sco, 16 days after outburst, when the SSS phase had already ended \citep{Drake2016}. There is also information about X-ray emission during quiescence: between the two last outbursts, V745 Sco was marginally detected with \textit{XMM-Newton} in the energy range 0.3-8 keV, indicating the existence of a hot plasma \citep{LunaATel2014}.

\subsection{Early X-ray observations with Swift}\label{sec:early_observations}

We present the analysis of V745 Sco  \textit{Swift}/XRT observations between days 0.16 and 4.23 after the eruption,  before the start of the SSS phase, and on day 16.03 post eruption, after the SSS phase ended and contemporaneous with the \textit{Chandra} observation \citep{Drake2016}. These early observations by \textit{Swift}/XRT coincide with the epoch of the expected particle acceleration in the shock, 1 to 4 days after the outburst  \citep{CheungATel2014}, thus enabling us to study the hot plasma  properties. Most of the observations were made before the nova declined by 3 magnitudes, and they showed a harder spectrum  than expected for the supersoft phase (which reveals residual H burning on top of the WD); this emission originates in the hot shocked ejecta. The log of the \textit{Swift}/XRT observations is shown in Table \ref{tab:observaciones}. These observations began 3.7 hours after the eruption and continued with a high cadence (6 hours) giving a very good temporal coverage.

\begin{table}

\caption{Observation log of V745 Sco with \textit{Swift}/XRT}
\label{tab:observaciones}
\scalebox{0.8}{\begin{tabular}{ l c c c c }
\hline
\hline
   				&  &Time after  	& Exp.time 	& Count rate\\
Observation Date	& ID &discovery		& (ks)		& (count s$^{-1}$)\\
\hline
\hline\\
 2014 Feb 06 	&00033136001&0.16					&0.98	& 0.28$\pm$0.01 	\\
 2014 Feb 07 	&00033136002&0.56					&0.98	& 0.35$\pm$0.01 	\\
 2014 Feb 07 	&00033136003&0.89					&0.91	& 0.37$\pm$0.02 	\\ 
 2014 Feb 07 	&00033136004&1.16					&1.77	& 0.40$\pm$0.01 	\\
 2014 Feb 08 	&00033136005&1.51					&0.89   	& 0.46$\pm$0.02	\\
 2014 Feb 08 	&00033136006&1.89					&0.96   	& 0.58$\pm$0.02	\\
 2014 Feb 08 	&00033136007&2.17					&0.98   & 0.49$\pm$0.02	\\
 2014 Feb 09 	&00033136012&2.96					&0.97  	& 0.54$\pm$0.02	\\
 2014 Feb 10 	&00033136017&4.23					&1.03 	& 10.6$\pm$0.1 	\\
 2014 Feb 22$^{(a)}$		&00033136056&16.03			&1.47	& 0.41$\pm$0.02 \\
\hline
\hline
\end{tabular}}

(a) After super soft emission.

\end{table}

\subsubsection{Data Reduction} \label{sec:data_reduction}

Data were reduced using HEASoft (version 6.15). We used standard routines of FTOOLs for the extraction of spectra. The CCD detectors of the XRT instrument have three different modes: IM, WT mode and PC mode. The first days, novae are very bright and usually there is pile-up. This effect causes the camera detectors to get an energy that is twice (or more) the correct energy, and thus the  highest energies of the spectra are contaminated.  To reduce the pile-up in PC spectra,  we use an annular extraction region with an inner radius between 2.5 to 5 pixels (1 pixel = 2.36$"$). The same area is extracted away from the source to obtain a background spectrum.  On day 4.23, the  SSS emission was detected using WT mode. The WT mode is the best mode to reduce the effects produced by the pile-up because for this mode pile-up only appears for count rates higher than 30 count s$^{-1}$. 

\subsubsection{Data Analysis} \label{sec:data_analysis} 

\begin{figure*}

	\includegraphics[angle=-90,width=1.5\columnwidth]{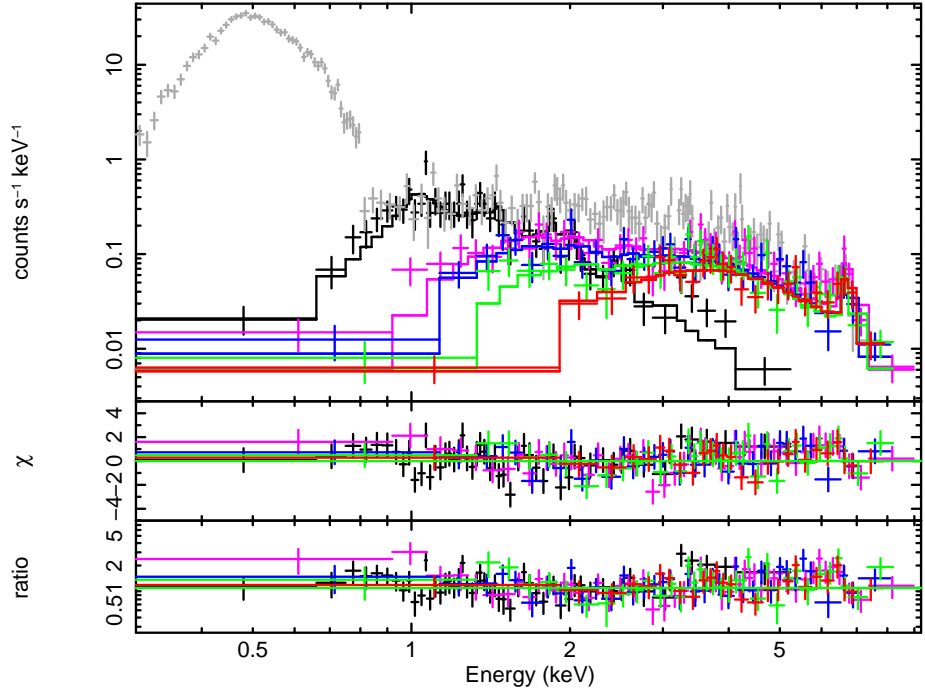}
    \caption{\textit{Swift}/XRT observed spectra and best-fit models for V745 Sco at days 0.16 (red), 0.89 (green), 1.51 (blue), 1.89 (pink) and 16.03 (black), and the SSS phase spectrum for day 4.23 (gray). Residuals are plotted in the bottom panels. }
    \label{swift}
\end{figure*}

We study the early hard X-ray spectra obtained with \textit{Swift}/XRT using XSPEC \citep[version 12.8,][]{Arnaud2012}. V745 Sco is a system very similar to RS Oph and, as in that case, we expect the X-ray emission to be produced by the shock between the ejecta and the RG wind. The X-ray spectral evolution corresponding to the first days is shown in Figure \ref{swift}. 

We fit the XRT spectra with an absorbed single-temperature  APEC model \citep{Smith2001}.  APEC models include both line and continuum emission for an optically thin plasma in collisional ionization equilibrium. We always use tbabs absorption \citep{Wilms2000}. We know that a single-temperature model does not reproduce all the details of the spectrum but gives a global view of the properties of the plasma during the first days. The Fe K line complex at 6.69 keV is the only strong emission line visible in the spectra because of the low resolution of the data.  Solar abundances  are used \citep{Anders1989}, except for Fe. 

\begin{table*}

	\begin{center}

	\centering

	\begin{scriptsize}
	\caption{ Parameters of the VAPEC Models of V745 Sco}
	
	\label{model_values_Swift}

	\begin{tabular}{ l c |c c c | c c c c c c}
	\hline
	\hline\\[-0.8em]

 	Day 			& N$_{\text{H}}$\tiny{${(a)}$}   	& \multicolumn{3}{|c|}{kT} 						& Fe/Fe$_{\odot}^{(b)}$ 	& EM $^{(a, d)}$ 					&F$_{\text{abs}}^{(b, e)}$&\multicolumn{2}{c}{F$_{\text{unabs}}^{(e)}$} 	& $\chi^{2}$/dof, $\chi^{2}_{red}$\\
 	   			& \tiny{($10^{22} cm^{-2}$)}	& \multicolumn{3}{|c|}{\tiny{(keV)}} 			&  						&\tiny{$\times$10$^{58}$cm$^{-3}$}	&  						& 							&  					& \\[0.2em]
     			& 							& \tiny{${(a)}$}&\tiny{${(b)}$} & \tiny{${(c)}$}	&  						& 									&       					& 	\tiny{${(b)}$}			&\tiny{${(c)}$} 		& \\

	\hline

	\hline\\[-0.6em]
 	0.16 	&5$^{+3}_{-2}$ 					&$>$5			&10$\pm$5		&[6 - 31]		& 2$\pm$1	   			&4$_{-2}^{+3}$						&  6$_{-1}^{+0.3}$  		& 10.5$_{-1}^{+0.5}$     	&[9 - 11]		& 18.7/23, 0.8 	\\[0.2em]
 	0.56		&3$\pm$2							&$>$6			&12$\pm$3		&[8 - 28]		& 2.2$\pm$0.6			&4$_{-3}^{+6}$						&  7.13$\pm$0.05  		& 11.36$_{-0.07}^{+0.08}$	&[11.3 - 11.4]	& 49.5/29, 1.7 	\\[0.2em]
 	0.89		&3$^{+2}_{-1}$	    				&$>$5			&9$\pm$3			&[6 - 24]		& 1.8$\pm$0.8			&5$_{-3}^{+4}$		    				&  8.27$\pm$0.04 		& 12.8$_{-0.6}^{+0.7}$ 		&[12 - 13]		& 43.9/32, 1.4 	\\[0.2em]
 	1.16		&2.0$^{+0.8}_{-0.5}$				&$>$7			&12$\pm$3		&[8 - 25]		& 2.5$\pm$0.9			&4$\pm$2								&  8.8$_{-0.9}^{+0.4}$ 	& 12.5$_{-1}^{+0.6}$ 		&[11 - 13]		& 83.1/67, 1.2 	\\[0.2em]
 	1.51		&1.7$^{+0.9}_{-0.6}$				&$>$5   			&12$\pm$3		&[7 - 26]		& 1.7$\pm$0.8			&4$_{-2}^{+3}$						&  7.8$_{-0.5}^{+0.4}$  	& 10.8$\pm$0.6				&[10 - 11]		& 37.1/35, 1.0	\\[0.2em]
 	1.89		&1.7$^{+0.7}_{-0.6}$				&$>$6   			&10$\pm$2		&[7 - 14]		& 2.7$\pm$0.8			&3$\pm$2								&  7.7$\pm$0.3 			& 10.5$\pm$0.4				&[10 - 11]		& 59.8/51, 1.2	\\[0.2em]
 	2.17		&1.7$\pm$0.5						&$>$6   			&9$\pm$2			&[6 - 16]		& 2.8$\pm$0.9			&4$\pm$2								&  9.7$\pm$0.4   		& 13.3$\pm$0.6				&[13 - 14]		& 52.8/44, 1.2	\\[0.2em]
 	2.96		&1.9$\pm$0.6						&$>$8  			&13$\pm$2		&[9 - 26]		& 2.8$\pm$0.9		   	&4$\pm$2								&  8.2$_{-0.5}^{+0.3}$  	& 11.4$_{-0.5}^{+0.6}$ 		&[11 - 12]		& 86.8/47, 1.8	\\[0.2em]
 	4.23	$^{(f)}$&1.2$^{+0.4}_{-0.5}$ 		&$>$5 			&8$\pm$1			&[7 - 12]		& 0.8$\pm$0.2			&6$\pm$3								&  10.2$\pm$0.3			& 13.9$\pm$0.4				&[13 - 14]		& 106.7/66, 1.6 \\[0.2em]
 	
	16.03$^{(g)}$&0.5$\pm$0.2					&1$^{+1}_{-0.2}$&1.23$\pm$0.06	&[1.1 - 1.3]		& 0.14$\pm$0.04 			&3$_{-2}^{+1}$		    				&  2.3$\pm$0.1			& 3.96$_{-0.2}^{+0.1}$ 		&[3.8 - 4.1]		& 66.6/49, 1.3 	\\[0.2em]
	\hline
	\hline
	\end{tabular}
 	\vskip 0.1 cm 
 	\hskip 0.5 cm
 	\begin{raggedright}
	
		 (a)  Uncertainties correspond to 3$\sigma$.\\
		 (b)  Uncertainties correspond to 1$\sigma$.\\
		 (c)  Uncertainties correspond to 2$\sigma$.\\
		 (d)  norm$_{VAPEC}$=10$^{-14}$EM/(4$\pi$D$^{2}$), where EM is the emission measure and D the distance in cm. EM=$\int_{\text{V}} $n$_{i}$ n$_{e}$ $d$V, where n$_{e}$ and n$_{i}$ are the electron and ion densities respectively.  We adopt a distance of 7.8$\pm$1.8 kpc ((2.4$\pm$0.5)$\times$10$^{22}$cm) \citep{Banerjee2014}. \\
		 (e)  0.8-10.0 keV. Parameter units are $\times$10$^{-11}$erg cm$^{-2}$s$^{-1}$.  	 \\
		 (f)  Fits to 1.0-10 keV, ignoring super soft emission.\\
		 (g)  After super soft emission turn-off.\\
        \end{raggedright} 
 	\end{scriptsize} 	
	\end{center}
	
\end{table*}

\begin{figure}
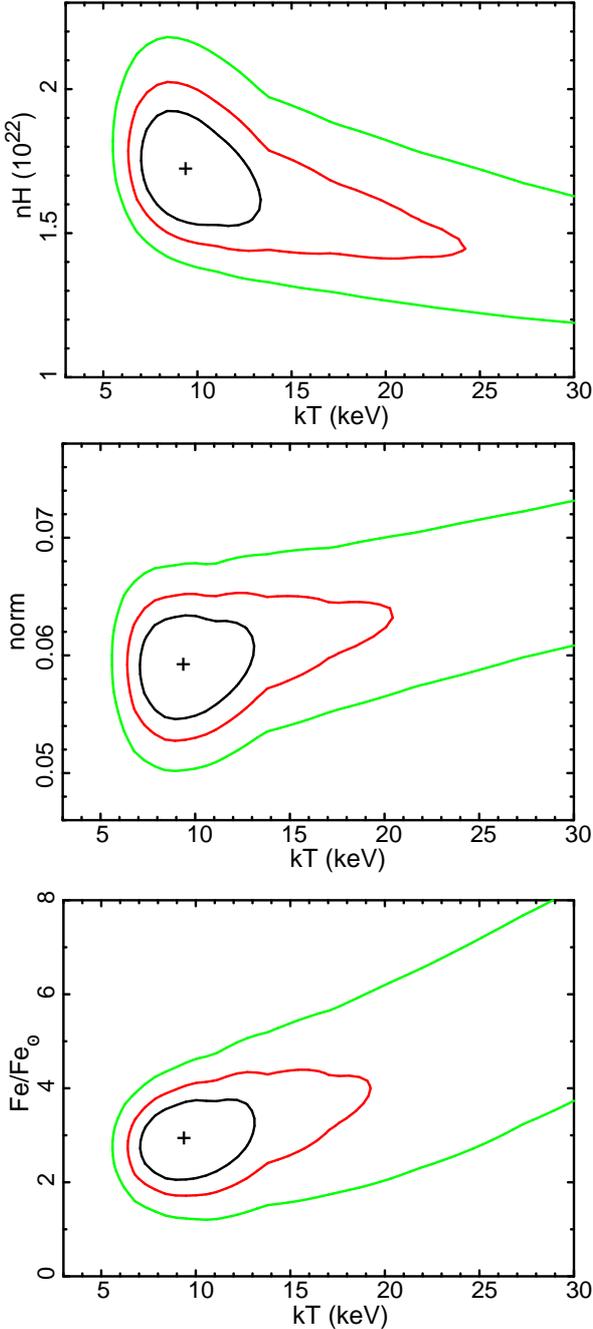

	\includegraphics[angle=-90, width=0.45\textwidth]{./kT_nH_50.ps}\\
	
	\includegraphics[angle=-90, width=0.45\textwidth]{./kT_norm_50.ps}\\
	
	\includegraphics[angle=-90, width=0.45\textwidth]{./kT_Fe_50.ps}
    \caption{Confidence contours (1 (black line), 2 (red line), and 3 $\sigma$ (green line)) of N$_{\text{H}}$-kT, normalization-kT and Fe/Fe$_{\odot}$-kT of  VAPEC model fitting the \textit{Swift}/XRT data  2.17 days after the outburst.  }
    \label{contour_XRT}
\end{figure}

We obtain Fe abundance around 2 times solar for the first days. It decreases to $\sim$ 0.1 Fe/Fe$_{\odot}$ on day 16.03 after the outburst, a similar  value to that from \citet{Drake2016}. Fe/Fe$_{\odot}$ errors are 2$\sigma$. If we do not modify the Fe abundances, a discrepancy between the data and the model is seen at energies lower than those of the broad Fe K complex. This excess can be interpreted as an additional line at 6.4 keV which is typical of cold (not ionized) Fe, as in some magnetic CVs, also  observed in the early \textit{Swift} spectra of RS Oph. In that case it was considered as scattering in the RG wind \citep{Bode2006}. 
In the \textit{NuSTAR} observation, 10 days after outburst \citep{Orio2015},  a good fit for the energy range between 3 and 20 keV was found with a Fe abundance of 0.51 Fe/Fe$_{\odot}$, but the uncertainty in the Fe abundance was not reported. We try to fit  the observations with a VAPEC model with the Fe abundances fixed to the same value of the \textit{NuSTAR} observation. Even with a good $\chi^2$, the optimization procedures of XSPEC only found a good fit at high temperatures, at which it is not possible to have a Fe K line at 6.69 keV. These high temperatures were shown in the study of the first days by \citet[][Table 1]{Page2015}.   It would have been interesting to know the uncertainties of the Fe abundances of \textit{NuSTAR} to know the true nature of the excess in the iron line. 

The high absorption N$_{\text{H}}$ during the first days of the nova outburst shows that there was not only the ISM absorption \citep[N$_{\text{H,ISM}}$=(5$\pm$1)$\times$10$^{21}$ cm$^{-2}$,][]{Page2015}, but also the absorption by the nova ejecta and the  RG wind. As the measured N$_{\text{H}}$ depends on the abundances used, we find discrepancies with \citet{Page2015}'s N$_{\text{H}}$.

The fits include the whole XRT data range (0.3 - 10 keV) until the emergence of the super soft emission on day 4.23 after the outburst. This day, we only use the data at energies higher than 1.0 keV. Table \ref{model_values_Swift} shows the parameters of the model used to fit the spectra. Confidence contours for the total hydrogen column density (N$_{\text{H}}$), temperature, Fe abundance and normalization factors are displayed in Figure \ref{contour_XRT} for day 2.17 after the outburst.

\subsection{NuStar and Chandra observations}

V745 Sco was also detected with \textit{NuStar} 10 days after its outburst. A detailed analysis of this observation is reported in \citet{Orio2015}.  During this observation, the supersoft emission was dominant but not observable in the energy range of \textit{NuSTAR} ($3-79$ keV). V745 Sco was detected in the energy range $3-20$ keV with a flux of 1.7$\pm$0.1$\times$10$^{-11}$ erg cm$^{-2}$ s$^{-1}$; above this energy, between 20-70 keV, only upper limits were obtained ($\sim$10$^{-13}$ erg cm$^{-2}$ s$^{-1}$). Taking into account the very fast evolution of V745 Sco, \textit{NuStar} observation was probably too late to detect emission at  energies larger than 20 keV, and it was thus not possible to detect the expected non-thermal emission related with the acceleration of particles. 
\textit{NuSTAR} clearly detected the iron He-like triplet at 6.73 keV with an equivalent width of 1 keV. \citet{Orio2015} fitted the spectrum  with a single temperature VAPEC model with kT=2.66 keV and Fe$/$Fe$_{\odot}$=0.51, but, as already mentioned, they did not provide the error for the abundance. 

Moreover, V745 Sco was detected with  \textit{Chandra} \citep{Drake2016} 16 days after its outburst, when the SSS phase had ended.  The \textit{Chandra}/LETG observation gave similar results to those of the \textit{Chandra}/HETG observations but with larger uncertainties. To study the absorption, \citet{Drake2016} checked a partial covering model, but finally found better fits using a single hydrogen column density. They tested  APEC models with a single temperature, three temperatures and eight temperatures using in all cases solar abundances \citep{Anders1989}. An APEC model with 8 temperatures gives an idea about the temperature distribution in the plasma, while a single temperature gives the dominant temperature in the corresponding energy range.  They also varied the abundances, the line broadening and the redshift of the lines, obtaining abundances lower than solar, and lines with a velocity broadening of approximately 510 km s$^{-1}$ and with a blueshift corresponding to 165$\pm$10 km s$^{-1}$.

\subsection{Results of the X-ray observations}\label{sec:globalview}
 
We use the plasma properties (temperature, emission measure (EM), N$_{\text{H}}$, flux) obtained by \citet{Drake2016}  (\textit{Chandra} observations) and \citet{Orio2015} (\textit{NuStar}$+$\textit{Swift} observations) the first two weeks after the outburst plus those derived from our analysis presented in section 3.1, to get the temporal evolution of the plasma. 

\begin{figure}
 
 \includegraphics[width=0.5\textwidth]{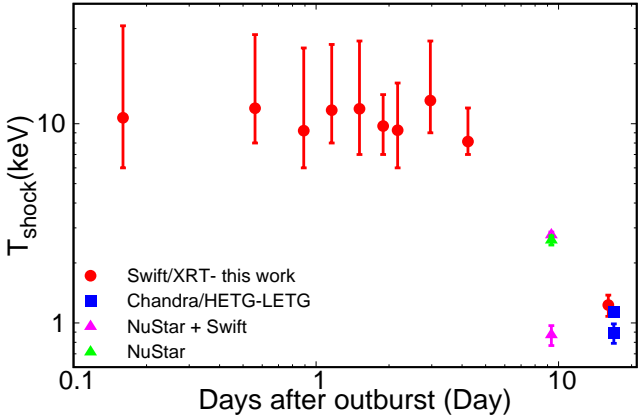}

\caption{Post-shock plasma temperature evolution with time. Red dots correspond to \textit{Swift}/XRT, blue boxes to \textit{Chandra}/HETG-LETG \citep{Drake2016}, green triangle to \textit{NuStar} and pink triangles  to \textit{NuStar} and \textit{Swift} together \citep{Orio2015}. \textit{Swift} measurements have 1$\sigma$ error bars (At 3$\sigma$, only lower limits are obtained). \textit{NuStar} and \textit{Chandra} values have 3$\sigma$ error bars.}
\label{Temperature_V745Sco}
\end{figure} 
\begin{figure}

\includegraphics[width=0.5\textwidth]{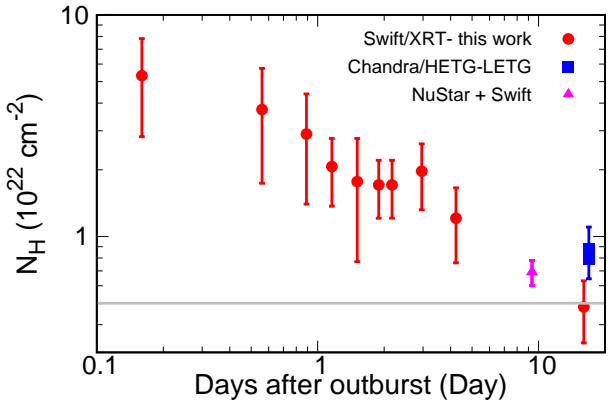} 
\caption{ N$_{\text{H}}$ evolution with time. Red dots correspond to \textit{Swift}/XRT,  blue boxes  to \textit{Chandra}/HETG-LETG \citep{Drake2016} and pink triangle to \textit{NuStar}+\textit{Swift} \citep{Orio2015}. All measurements have 3$\sigma$ error bars. Gray line corresponds to the ISM value.}
\label{NH_Swift_V745Sco}
\end{figure} 
 \begin{figure}

\includegraphics[width=0.5\textwidth]{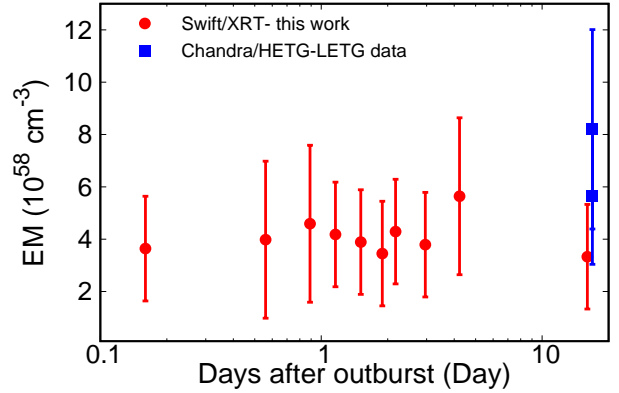} 
\includegraphics[width=0.5\textwidth]{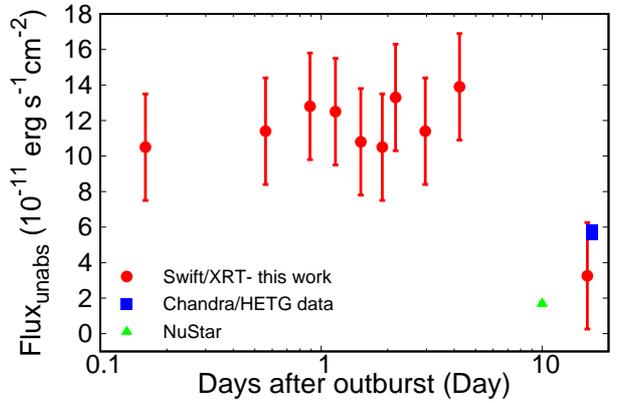} 

\caption{Top: Emission measure of the shocked plasma as a function of time. Bottom: Unabsorbed X-ray flux, F$_{\text{unabs}}$, as a function of time. Red dots correspond to \textit{Swift}/XRT (3$\sigma$ error bars),  green triangle to \textit{NuStar} \citep{Orio2015} and  blue boxes  to \textit{Chandra}/HETG-LETG (top) \citep{Drake2016} and HETG (bottom).  }
\label{EM_flux}

\end{figure} 

The temperature evolution is of great relevance for our study due to its relation with particle acceleration. Figure \ref{Temperature_V745Sco} shows the plasma temperature evolution obtained from the \textit{Swift}, \textit{NuStar}  and \textit{Chandra}/HETG-LETG  observations.  The temperature is practically constant the first days after the outburst, and it later decreases with time according to \textit{NuSTAR} and \textit{Chandra} observations. This decrease is faster than expected for an adiabatic shock due to radiative losses. In general, there is only one observation satellite at each epoch. It is worth noticing that the temperatures obtained from \textit{NuStar}  and \textit{Chandra}/HETG-LETG observations show the complexity of the plasma.  
                                                                                                                                                                                                                                                                                                                                                                                                                                   
Figure \ref{NH_Swift_V745Sco} shows the evolution of N$_{\text{H}}$. The spectrum is strongly absorbed three hours after the outburst, with X-ray counts only above 2 keV (Figure \ref{swift}). The absorption decreases very fast reaching the interstellar medium value 16 days after the eruption.    \citet{Bode2006} stated for RS Oph that all N$_{\text{H}}$ values greater than N$_{\text{H,ISM}}$ correspond to the absorption produced by the stellar wind of the RG (N$_{\text{H,W}}$), so  N$_{\text{H}}$= N$_{\text{H,W}}$ + N$_{\text{H,ISM}}$. In our case, we assume that this absorption is also related to the ejected material.    
 
The emission measure gives information about the mass  and particle (electron, ion) densities of the hot material that emits X-rays. The EM obtained by fitting the \textit{Swift}/XRT spectra is practically constant with time, even up to day 16 after the outburst, although the error bars are large (Figure \ref{EM_flux}). The EMs obtained from \textit{Chandra} data are compatible with \textit{Swift}/XRT within the error bars. The highest EM corresponds to \textit{Chandra}/LETG \citep{Drake2016}, which has a better sensitivity to lower energies than \textit{Swift}/XRT and \textit{Chandra}/HETG.
 
Finally, we  study the evolution of the unabsorbed X-ray flux (Figure \ref{EM_flux}), F$_{unabs}$, comparing \textit{Swift} F$_{unabs}$ [$0.8-10$ keV], \textit{Chandra} F$_{unabs}$ [$0.9-6.0$ keV]  \citep{Drake2016} and  \textit{NuStar} F$_{unabs}$ [$3-10$ keV] \citep{Orio2015}. F$_{unabs}$ increased slightly during the first days, being practically constant until the SSS phase begins. During the SSS phase, F$_{unabs}$ obtained by \textit{Swift} 10 days after the outburst increased by two orders of magnitude  \citep[][Table 2]{Orio2015}. For our study, we only take into account the result of the hard X-ray emission obtained with \textit{NuStar}. At day 16 after the outburst F$_{unabs}$ decreased to $\sim$4$\times$10$^{-11}$ erg cm$^{-2}$ s$^{-1}$.

\section{IR observations} \label{sec:IRobservations}

\begin{figure}
			\includegraphics[width=0.5\textwidth]{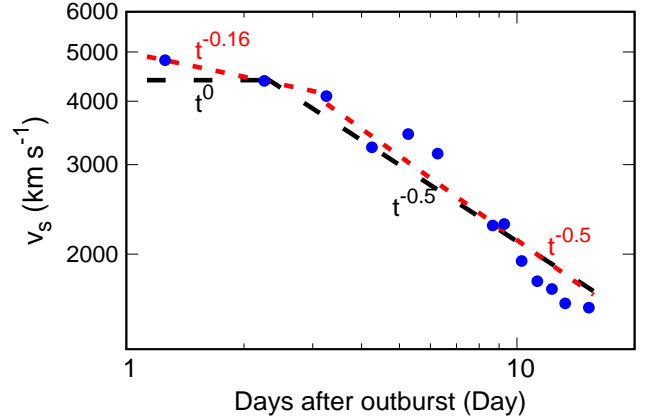}
			\caption{Temporal evolution of the broad gaussian line component FWHM of H I Paschen $\beta$ emission line between 1.3 and 15.3 days after the outburst \citep[][Figure 3]{Banerjee2014}. The velocity is fitted with two different models: in  the first model, the shocked plasma evolved from phase I to phase III on day 2  (black dashed line);  in the second one, the shocked plasma evolved from phase I$\rightarrow$II  to phase III on day 3 (red dashed line). }
			\label{fig:IRV745Sco_pendientes}
		\end{figure}

V745 Sco was detected in IR by \citet{Banerjee2014} the first days after the outburst. Observations started 1.3 days after the outburst, as for RS Oph, later than the first detection of X-rays.  \citet{Banerjee2014} measured the profile of the H I Paschen $\beta$ emission line between 1.3 and 15.3 days after the outburst, but they did not report  errors of their data.  H I Paschen $\beta$ was chosen because it is not blended with other lines. Observations were fit with two gaussians: a broad one related with the fast movement of the nova shell, and a narrow one corresponding to the slow movement of wind material, with constant velocity. These two components were also seen by  \citet{Duerbeck1989} in the previous eruption in 1989 and by \citet{AnupamaATel2014} in optical in the 2014 eruption. As we mention in Section 1, the near-IR emission originates in the contact surface between the ejected material and the RG wind.  The kinematic broadening, Full Width at Half Maximum (FWHM) and Full Width at Zero Intensity (FWZI), are good indicators of the shock velocity and FWZI/2 can be used as a good measure of the expansion velocity \citep{Das2006}.  We use the FWHM of the broad gaussian line component  given by \citet{Banerjee2014} as a measure of the forward shock velocity (v$_{s}$) of V745 Sco.  In Figure  \ref{fig:IRV745Sco_pendientes} , we  show how v$_{s}$ decreases very slowly during the first 4 days. After day 4, the slope becomes steeper and the forward shock velocity is drastically reduced, which indicates that the expanding material is now braked by the circumstellar medium. Thanks to these observations,  we estimate v$_{s}$  the first days after the outburst. We also know that the FWZI of Pa$\beta$ was 9130 km/s when the FWHM was approximately 4000 km/s \citep{Banerjee2014ATel}. The forward shock velocity is approximately  FWZI/2,  slightly larger than the FWHM. For this reason we probably are underestimating v$_{s}$. 

Novae behave as miniature supernova remnants (SNR). They are much dimmer and evolve much faster than SNR. Therefore,  we can use the characteristics of the SNR evolution  to understand the phases of the nova blast wave, scaling them down to a much shorter period of time. In the case of symbiotic novae, whose companion is a RG, the system is surrounded by a circumstellar density distribution of matter $\rho\propto r^-2$. Therefore, we use the models presented by \citet{Chevalier1982, ChevalierL1982} and \citet{Truelove1999} to obtain the phases of the shock. \citet[][Figure 3]{Banerjee2014} fitted the IR data with the power law associated to the adiabatic phase (v $\propto t^{-1/3}$), but this was clearly not a good fit to the data.  They found that the best fit was a third degree polynomial. However, a polynomial does not correspond to any phase in the evolution of the shocked plasma. What we do is to fit the data taking into account the different evolutionary phases of the nova ejecta, revealed by the different behaviours of the shock speed, using two power laws. It is important to keep in mind that we do not know the errors in the measure of the FWHM, so that we fit the data directly but knowing that the error could change the result.

\begin{itemize}

\item \textbf{First fit} (black dashed line).  The first epoch starts on day 1.3 after the outburst with a constant velocity ($\alpha$=0) and the second one starts on day 2 after the outburst with a power law index of $\alpha$=-0.5. These indexes correspond to Phase I (v = const.)  \citep{Woltjer1972}  and Phase III (v $\propto t^{-1/2}$)  \citep{Truelove1999} of SNR evolution. The ejected mass expanded freely through the circumstellar medium until day 2, when it started to lose momentum by radiation, skipping the adiabatic phase. 

\item \textbf{Second fit} (red dashed line). In this model, we assume that the change in the slope occurs on day 3 after the outburst. We fit the data with a power law index of $\alpha$=-0.16 the first days after the outburst, and a power law index of $\alpha$=-0.5 in the second time interval. This corresponds to Phase I$\rightarrow$II (v $\propto t^{-1/(n-2)}$ where n=8) \citep{Chevalier1982, ChevalierL1982, Chevalier21982} and Phase III (v $\propto t^{-1/2}$) \citep{Truelove1999}. This model would indicate that the shocked plasma was in a transition between the free expansion and the adiabatic phase and ended up losing momentum by radiation directly, without reaching the adiabatic phase.

\end{itemize} 

Other combinations are possible, but without knowing the errors there are open questions, such as when the slope change takes place, what is the state of the system in the first days, and also what is its state during the first hours, for which we do not have IR data. We conclude that the shocked plasma is in the radiative phase the last days and that it skipped the adiabatic phase. The reason is most probably that the nova cooled faster than expected. According to \citet{Tatischeff2007}, this may be due to the diffusive acceleration of particles in a collisionless shock \citep{Jones1991}. With the aim of better understanding the properties of V745 Sco during the first two weeks after outburst, we combine IR data with data at other wavelengths, to get a global picture of the V745 Sco evolution.

\section[Properties and evolution of the plasma]{Properties and evolution of the plasma  behind the forward shock} \label{shockmaterial}

We have obtained information about the shock wave expanding into the RG wind using observational data at several wavelengths. X-ray observations provide the most representative temperature of the shock, whereas IR observations give  the velocity of the shock. We use v$_{s}$ obtained from IR data as the velocity of the forward shock. IR observations were not simultaneous to \textit{Swift}/XRT  observations; they started 1 day after the outburst, so there is no information about the behaviour of v$_{s}$ during the first hours after outburst. We analyse the different properties from day 1 after outburst and we interpolate the v$_{s}$ corresponding to the \textit{Swift}/XRT,  \textit{NuStar} and \textit{Chandra} observations times using the FWHM of the broad gaussian line component of the H I Paschen $\beta$.  Another approach employed by some authors is to assume that v$_{s}$ is constant with time. However, it is more appropriate to take into account the temporal evolution of v$_{s}$ to derive the properties of the plasma, as we will show in this section.

With v$_{s}$ at the X-rays and IR observations times, we calculate the forward shock radius  (r$_{s}$) using 
\begin{equation}
r_{s}(t)=\int^{t_f}_{t_0} v_{s}(t) dt\longrightarrow r_{s}(t_{i+1})= r_{s}(t_{i})+v_{s}(t_{i})\Delta t
\label{eq:radio}
\end{equation} 

There are some relevant dimensions in the geometry of the system. First of all, we have obtained the separation between the WD and the RG from Kepler's third law. Then we assume M$_{\text{RG}}$ $\sim$1 M$_{\odot}$: \citet{Banerjee2014} state that the M giants of S-type systems have a  low-mass ($<$1 M$_{\odot}$). Since V745 Sco is a recurrent nova, we adopt a massive WD (M$_{\text{WD}}$ 
$\sim$1.2 -1.4  M$_{\odot}$). As  explained in $\S$\ref{sec:V745Sco}, the orbital period of the system is not well known; it ranges between 77 and 255 days. Therefore, the separation between the WD and the RG ranges between $\sim$0.65 AU and $\sim$1.6 AU.  Another important dimension is the outer radius of the RG wind. The time between the two last outbursts of V745 Sco is $\Delta t=$25 yr, so that  r$_{\text{out}}$ ranges between 7.8$\times$10$^{14}$ cm and 1.6$\times$10$^{15}$ cm depending on u$_{\text{RG}} (\cong$10-20 km s$^{-1}$). 
 
\begin{figure}
\includegraphics[width=0.5\textwidth]{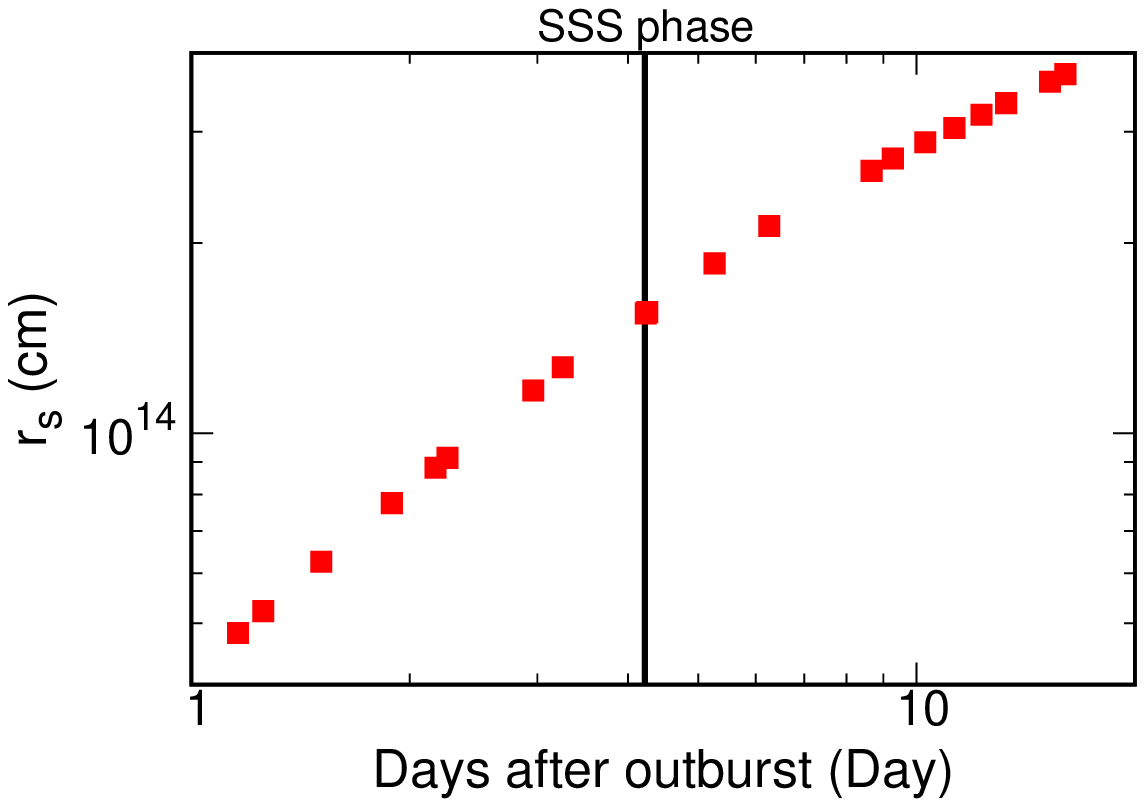}
\includegraphics[width=0.5\textwidth]{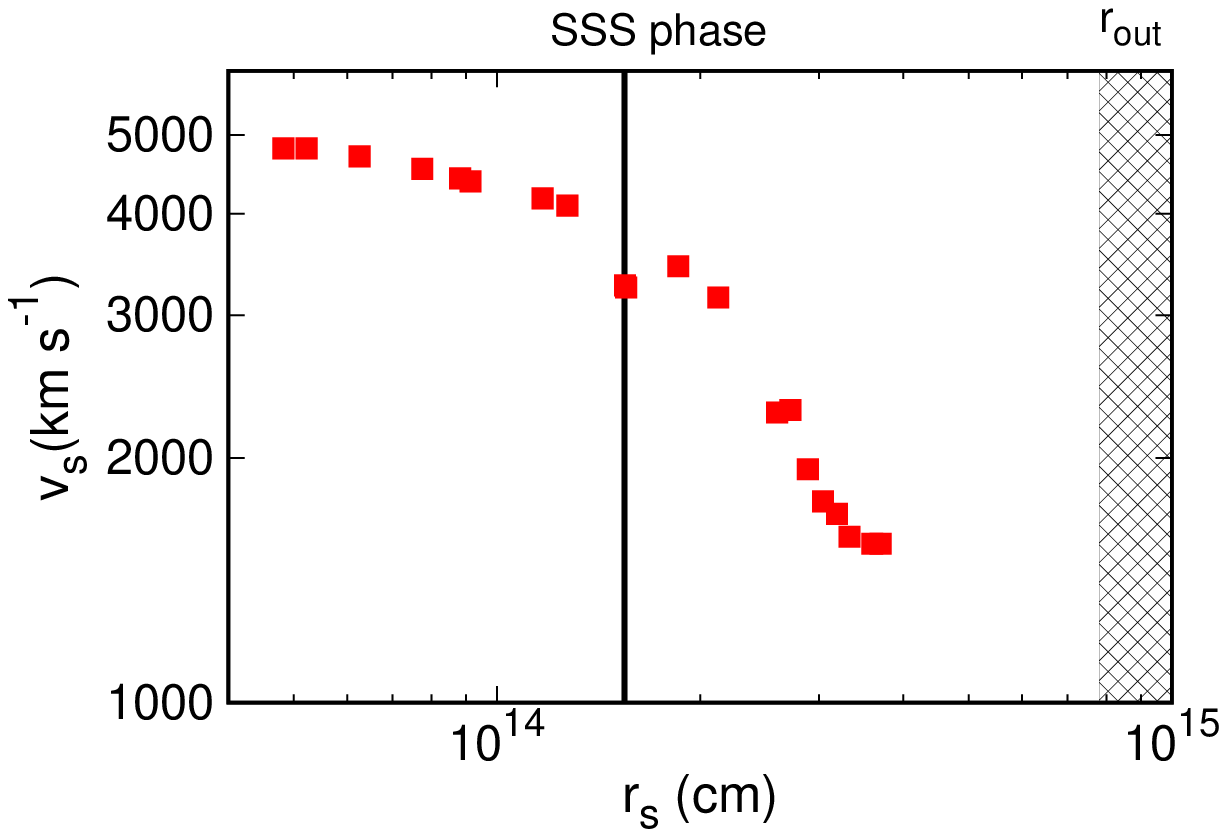}
\caption{Top: Evolution of forward shock radius with time. Bottom: Forward shock velocity versus forward shock radius. Shaded areas show the location of the shock when the SSS phase begins and r$_{\text{out}}$.  }
\label{fig:R_eje_models}
\end{figure}	

Figure \ref{fig:R_eje_models} shows the temporal evolution of the forward shock radius.  The  plasma behind the forward shock reaches the limit of the binary system in the early hours before the first IR observation.  We assume that the slope (the shock speed) does not change during this early epoch, but it is possible that during the first hours the expansion of the ejecta is homologous (v$_{s} \propto$ r$_{s}$) before the free expansion phase. If even earlier IR observations were available, it would be possible to clarify the evolution of the forward shock radius and velocity during the first hours, but this is observationally challenging.  We see a change in the slope of v$_{s}$  simultaneous to the appearance of the SSS phase (r$_{\text{SSS}}\sim$1.55$\times$10$^{14}$ cm). It is noteworthy to mention that r$_{s}$ are smaller than r$_{\text{out}}$ during the observations of our study. At r$_{\text{out}}$, it should happen that N$_{\text{H}}$=N$_{\text{H,ISM}}$, but we obtain N$_{\text{H}}$=N$_{\text{H,ISM}}$ before reaching this radius (see Figure \ref{NH_Swift_V745Sco}).  

The volume of the material behind the forward shock (Vol$_{s}$) is given by
\begin{multline}\label{eq:Volume}
\text{Vol}_{s}(t)= \text{Vol}_{s}(t_{0})+\int^{t_{f}}_{t_{0}} 4 \pi r^{2}_{s}(t)\cdot v_{s}(t) dt \longrightarrow \\
\text{Vol}_{s}(t_{i+1}) = \text{Vol}_{s}(t_{i}) + 4 \pi r^{2}_{s}(t_{i})\cdot v_{s}(t_{i}) \Delta t
\end{multline}

Since the location of the reverse shock is not known for V745 Sco, we can't obtain the volume of the region between the forward and the reverse shocks. Therefore, eq. \ref{eq:Volume} gives an upper limit to the shocked plasma volume.

\begin{figure}
\includegraphics[width=0.5\textwidth]{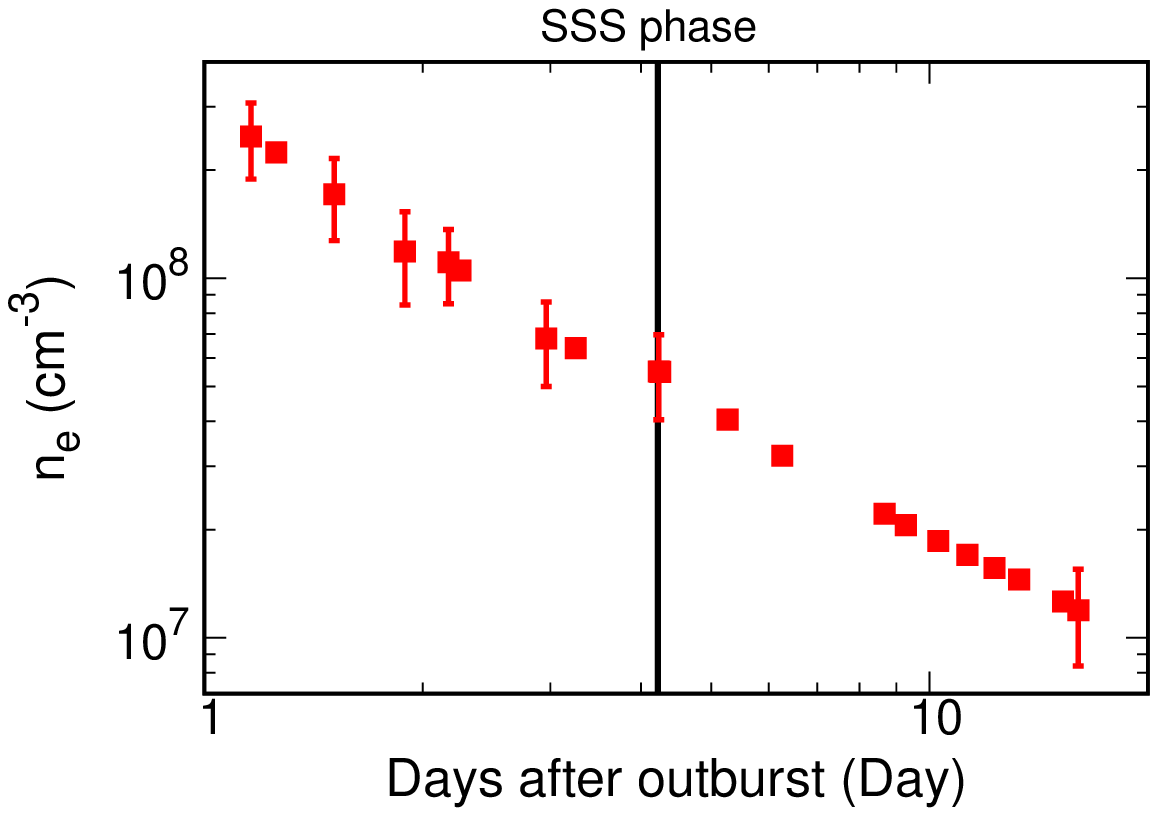}
\includegraphics[width=0.5\textwidth]{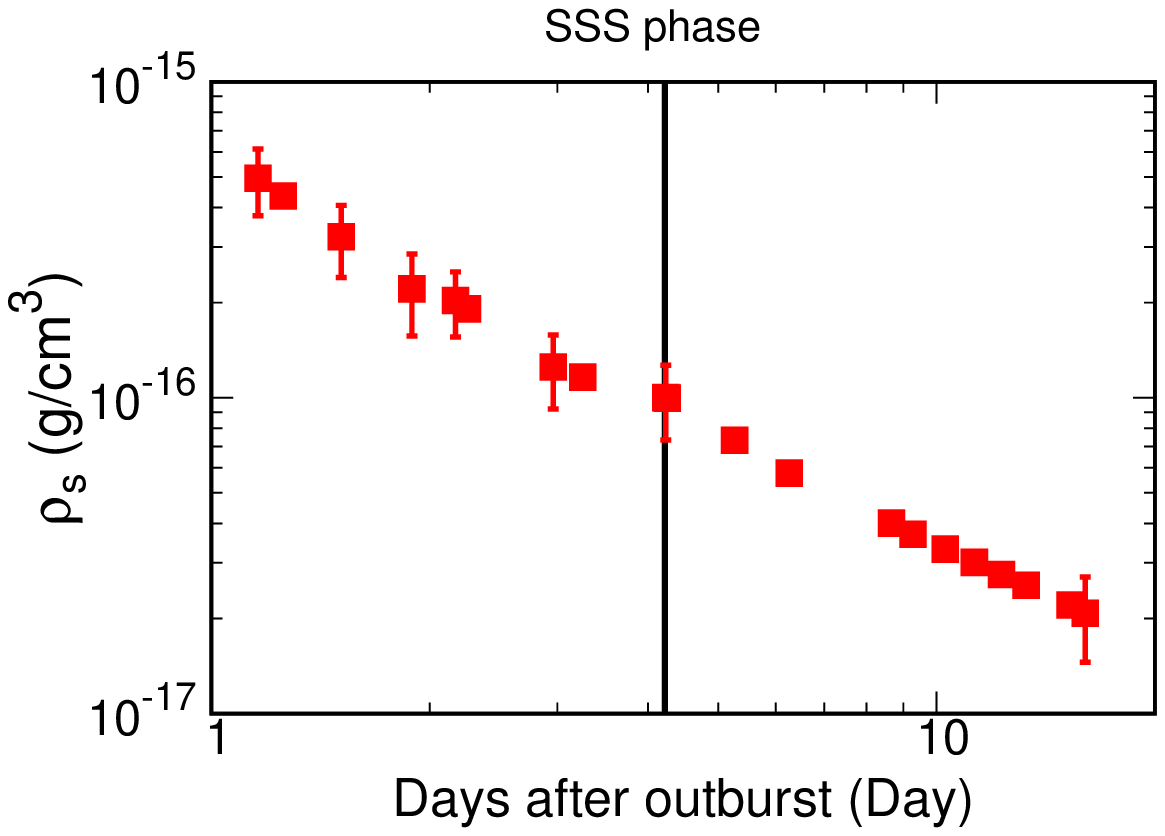}
\includegraphics[width=0.5\textwidth]{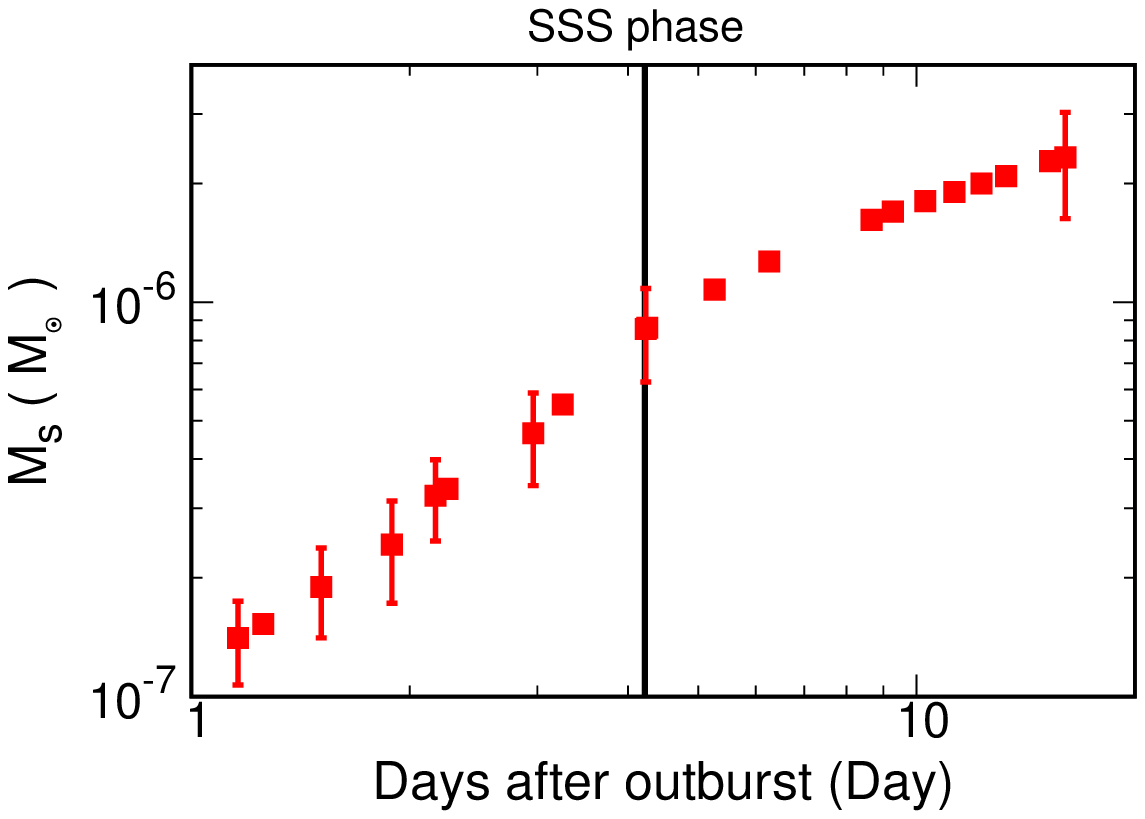}

\caption{Top: Evolution of the average electron density (lower limits). Middle: Same for the average mass density (lower limits). Bottom: Same for the  mass (upper limits) behind the forward shock of V745 Sco. The vertical line shows when the SSS phase begins.}
\label{fig:Vol_ne_rho_M_models}
\end{figure}

The evolution of the EM  is obtained for the most representative temperatures of the shocked plasma (Figure \ref{EM_flux}). But the value of the EM can't be deduced for epochs with IR detections but without X-ray observations; we extrapolate the EM for these times. Knowing the EM and the volume, we can estimate the average electron density (n$_{e}$), for which we can only obtain lower limits.  Figure \ref{fig:Vol_ne_rho_M_models} shows that  n$_{e}$ decreases faster the first four days and slower afterwards, reaching $\sim$10$^{7}$ cm$^{-3}$ 16 days after outburst.

The average mass density of the plasma behind the forward shock is given by $\rho_{s}=n_{e}\times \mu_{e} / N_{a}$ where $\mu_{e} =2/(X+1)$ and $X$ is the H mass fraction. It follows the same behaviour as the average electron density, reaching values of  $\sim$10$^{-17}$ g$/$cm$^{-3}$ 16 days after the outburst (Figure \ref{fig:Vol_ne_rho_M_models}). In novae closer than V745 Sco, as RS Oph \citep{Nelson2008, Ness2009} and V959 Mon \citep{Peretz2016} additional information about densities can be obtained from the H- and He-like transitions. \citet{Peretz2016} observed some clusters in the ejecta thanks to their density study. Unfortunately,  Swift observations of V745 Sco don't give any information about H- and He-like transition lines, so that only an average of the density can be deduced.

The mass behind the forward shock  is defined by M$_{s}$=$\rho_{s}\times$Vol$_{s}$. Figure \ref{fig:Vol_ne_rho_M_models} shows how mass increases as a function of time. M$_{s}$ includes the ejected mass (M$_{\text{ej}}$) and the swept-up material (M$_{\text{sw}}$).  We can only obtain upper limits of the mass behind the forward shock because we only have information about the upper limit of the shocked plasma volume.
We obtain two different behaviours for M$_{s}$: after 1 day, M$_{s}$ rapidly increases reaching 8$\times$10$^{-7}$ M$_{\odot}$ on day 4 after outburst, probably due to the increase of the M$_{\text{sw}}$. After day 8 M$_{s}$  becomes practically constant, because v$_{s}$ drastically decreases. Finally, M$_{s}$ reaches 2$\times$10$^{-6}$ M$_{\odot}$, 12 days after the beginning of the SSS phase.

\section{Properties and evolution of the red giand wind}  \label{sec:Redgiantwind}

The RG wind mass-loss rate, $\dot{\text{M}}_{\text{RG}}$, is required to determine the mass density of the RG wind ($\rho_{w}$) involved in the shock, and also to know the amplification of the magnetic field due to the acceleration of particles (B$_{0}$). These magnitudes give an idea of the power of the shock. To obtain the magnetic field associated to the red giant wind  (B$_{w}$), we need to find the density of the red giant wind ($\rho$), defined as:
\begin{equation}
\rho(t)=\dfrac{\dot{M}}{4\pi r^{2}v}\rightarrow \rho_{w}=\dfrac{\dot{M}_{\text{RG}}}{4\pi r^{2}u_{\text{RG}}}
\label{eq:rho}
\end{equation}
where u$_{\text{RG}}$ = 10 - 20 km s$^{-1}$ is the RG wind speed. In addition, the average density can be derived from the column density N$_{\text{H}}$:
\begin{equation}
\overline{\rho}(t)=\dfrac{N_{\text{H}} m_{H}}{hX}
\label{eq:rhopromedio}
\end{equation}
where $h$ is the height of an imaginary cylinder with a 1 cm$^{2}$ base  (see Figure 9). We adopt $X$=0.7, which corresponds to a solar-like composition. Next we will determine $\dot{\text{M}}_{\text{RG}}$ for V745 Sco, interesting by itself to characterise the RG, and needed to obtain the magnetic field and the mass density of the wind, B$_{w}$ and $\rho_{w}$.

\subsection{Determination of $\dot{M}_{\text{RG}}$}

\begin{figure}
\includegraphics[width=0.5\textwidth]{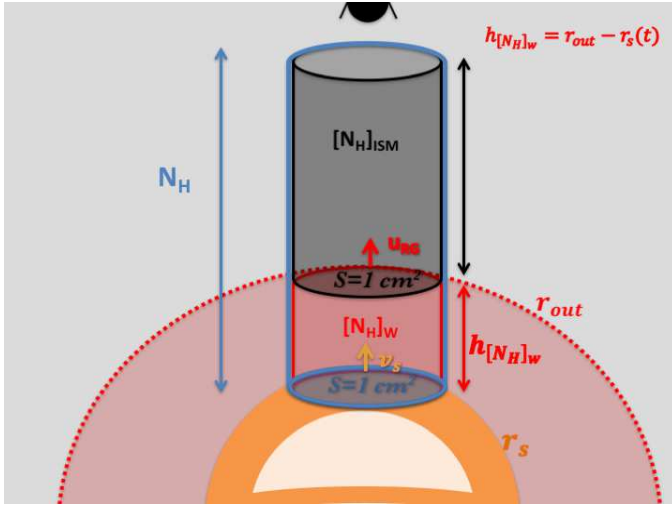}
\caption{Schematic view of the absorption N$_{\text{H}}$ and its components.}
\label{fig:Schema}
\end{figure}

In a nova explosion with a RG companion, there are three different components of N$_{\text{H}}$ in X-rays, the N$_{\text{H}}$ of the ejecta ([N$_{\text{H}}]_{\text{ejecta}}$), which is influential during the first few days, the N$_{\text{H}}$ of the red giant wind ([N$_{\text{H}}]_{w}$) and the N$_{\text{H}}$ of the interstellar medium ([N$_{\text{H}}]_{\text{ISM}}$). The material that we want to study is between two boundaries, only when [N$_{\text{H}}]_{\text{ejecta}}$ is negligible: the shocked plasma boundary that has radius r$_{s}$ and speed v$_{s}$, and the interstellar medium boundary that has radius r$_{\text{out}}$ and speed u$_{\text{RG}}$. Therefore, the height of a cylinder of  base with s=1 cm$^{2}$, $h$, is affected not only by the red giant wind speed, but also by v$_{s}$.  The absorption by the red giant wind is given by [N$_{\text{H}}]_{w}$= N$_{\text{H}}$-[N$_{\text{H}}]_{\text{ISM}}$ (see Figure \ref{fig:Schema}). 

The average radial density of the wind, $\overline{\rho_{w}}$(t), in the region between r$_{s}$(t) and r$_{\text{out}}$ with v=u$_{\text{RG}}$ is: 
\begin{multline}
\overline{\rho_{w}}(t)_{(r_{s}(t)\rightarrow r_{out})}=\dfrac{1}{r_{out}-r_{s}(t)}\int_{r_{s}(t)}^{r_{out}} \rho(r(t)) dr =\\
=\dfrac{\dot{M}_{\text{RG}}(t)}{4\pi u_{\text{RG}}(t)} \dfrac{1}{r_{out}r_{s}(t)}
\label{eq:rhopromediowind_1}
\end{multline}
where we have used Equation \ref{eq:rho}. From Equation \ref{eq:rhopromedio}, we get
\begin{equation}
\overline{\rho_{w}}(t)_{(r_{s}(t)\rightarrow r_{out})}=\dfrac{[N_{\text{H}}]_{w}(t) m_{H}}{h_{[N_{\text{H}}]_{w}} X}=\dfrac{[N_{\text{H}}]_{w}(t) m_{H}}{(r_{out}-r_{s}(t)) X}
\label{eq:rhopromediowind_2}
\end{equation}
 
Using Equation \ref{eq:rhopromediowind_1} and Equation \ref{eq:rhopromediowind_2}, we obtain:
\begin{equation}
\overline{\rho_{w}}(t)=\dfrac{\dot{M}_{\text{RG}}}{4\pi u_{\text{RG}}} \dfrac{1}{r_{out}r_{s}(t)}=\dfrac{[N_{\text{H}}]_{w}(t) m_{H}}{(r_{out}-r_{s}(t)) X}
\end{equation}  

and therefore [N$_{\text{H}}]_{w}(t)$ is given by 
\begin{equation}
[N_{\text{H}}]_{w}(t)=\dfrac{X}{4\pi m_{H}}\dfrac{\dot{M}_{\text{RG}}}{u_{\text{RG}}}\left( \dfrac{1}{r_{s}(t)}-\dfrac{1}{r_{out}} \right).
\label{eq:nH_w}
\end{equation}

The mass-loss rate $\dot{\text{M}}_{\text{RG}}$  of V745 Sco is not known. We assume that the red giant wind has $\dot{\text{M}}_{\text{RG}}$= const. and u$_{\text{RG}}$= const. We use Equation \ref{eq:nH_w}  to obtain a $\dot{\text{M}}_{\text{RG}}/$u$_{\text{RG}}$ for the different days under study.
We find three problems: we do not know for which days Equation \ref{eq:nH_w} is valid, N$_{\text{H}}$ shows large uncertainties, and we need to determine r$_{\text{out}}$. Figure \ref{NH_Swift_V745Sco} shows that N$_{\text{H}}$=[N$_{\text{H}}]_{\text{ISM}}$ at r$_{\text{out}}$=r$_{s}$(t=16 days)=3.7$\times$10$^{14}$ cm, that is smaller than r$_{\text{out}}=$7.8$\times$10$^{14}$ cm, as we mentioned in $\S$\ref{sec:globalview}. However, according to the \textit{Chandra} data, N$_{\text{H}} \neq $[N$_{\text{H}}]_{\text{ISM}}$ at r$_{\text{out}}$=r$_{s}$(t=16 days)=3.7$\times$10$^{14}$ cm. To tackle these problems we first carry out several tests. 
\begin{enumerate}
\item We apply Equation \ref{eq:nH_w} globally, dropping/adding one data point in each test, to accurately identify the radii at which the equation is satisfied. 
\item We use two different uncertainties, 3$\sigma$ error bars  and 1$\sigma$ error bars. 
\item We use two different values of r$_{\text{out}}$, which we assume constant for each one of the tests,   r$_{\text{out}}$=r$_{s}$(t=16 days)=3.7$\times$10$^{14}$ cm and r$_{\text{out}}=$7.8$\times$10$^{14}$ cm. 
\end{enumerate}
We find that using Equation \ref{eq:nH_w} globally,  $\dot{\text{M}}_{\text{RG}}/$u$_{\text{RG}}$ is not constant in time. There is not a single value of  r$_{\text{out}}$ that solves Equation \ref{eq:nH_w} for 
each one of the days. In most of the cases, we obtain a good approximation  from day 3 onwards (r$_{s}>$10$^{14} \text{ cm}$), finding better results when we apply r$_{\text{out}}$=r$_{s}$(t=16 days)=3.7$\times$10$^{14}$ cm in Equation \ref{eq:nH_w}. We conclude that the 3$^{\text{rd}}$ day after the outburst, the absorption of the red giant wind started to be dominant. Then,  we obtain $\dot{\text{M}}_{\text{RG}}/$u$_{\text{RG}}$ = (5 $\pm$ 1)$\times$10$^{13} \text{ g cm}^{-1}$ ($\dot{M}_{\text{RG}}$=[5 - 10]$\times$10$^{-7} $M$_{\odot}$ year$^{-1}$), which is similar to the value estimated by \citet{Tatischeff2007} for RS Oph. 

One of the problems that we found in this nova is that N$_{\text{H}}$ reached the value of $[N_{\text{H}}]_{\text{ISM}}$ earlier than expected. As a consequence of that, r$_{\text{out}}$ = r$_{s}$(t=16 days) works better in the models than r$_{\text{out}} = $u$_{\text{RG}}\Delta t$. We have three possible interpretations to explain this fact. The first and most likely is that the circumstellar medium becomes transparent earlier than expected because of photoionization. \citet{Nielsen2015} stated that photoionization of the wind material by the emission from the supersoft source can significantly reduce its opacity. In V745 Sco the SSS phase ended 4 days before N$_{\text{H}}$ reached the ISM value. 
The second is the possible non-spherical explosion of V745 Sco.  The asymmetries can cause the red giant wind to be swept-up faster than in spherical explosions, since the binary separation causes  a geometric offset between the spatial origin of the wind and that of the nova ejecta. The symmetry of the system was extensively discussed in \citet{Drake2016}. The third interpretation is that the recurrence period of V745 Sco is shorter than expected. r$_{\text{out}}$ depends on the time elapsed between the last two outbursts, which could be shorter than the generally adopted recurrence period if an intermediate outburst was missed. 
We calculate the time interval between the last two outbursts giving r$_{\text{out}}$=r$_{s}$(t=16 days). For u$_{\text{RG}}$=10 km s$^{-1}$, it would be $\sim$12 years, approximately half of the 
generally adopted recurrence period. For u$_{\text{RG}}$=20 km s$^{-1}$, the recurrence period would be 6 years. This value is the least probable, because nowadays recurrent novae are monitored almost  
continuously; however, since this nova is very fast and faint, it is not discarded. Other possibilities are that the RG wind is slower or that $[N_{\text{H}}]_{\text{ISM}}$ is underestimated. A study of the system during  quiescence would be necessary to accurately determine these two quantities and their possible evolution with time.

\subsection{Determination of $\rho_{w}$ and B$_{w}$}\label{sec:Redgiantwind}

\begin{figure}
\includegraphics[width=0.5\textwidth]{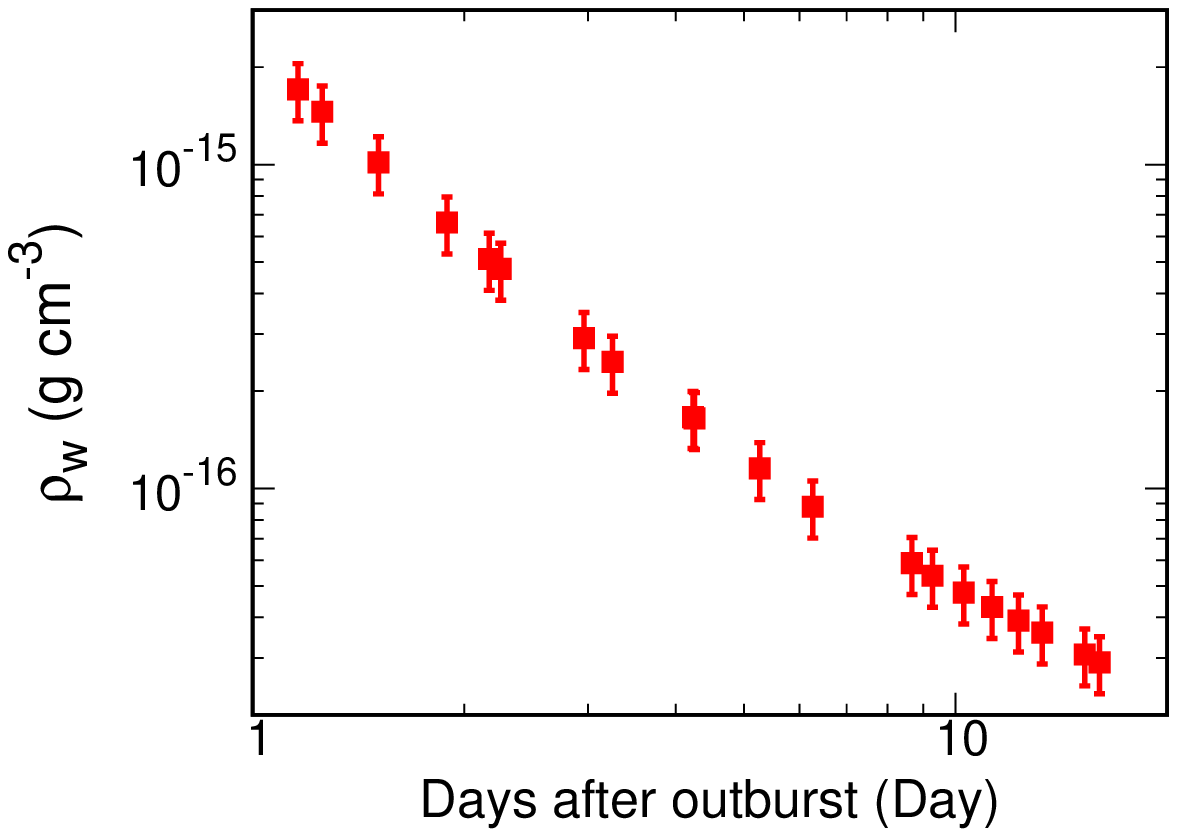}
\includegraphics[width=0.5\textwidth]{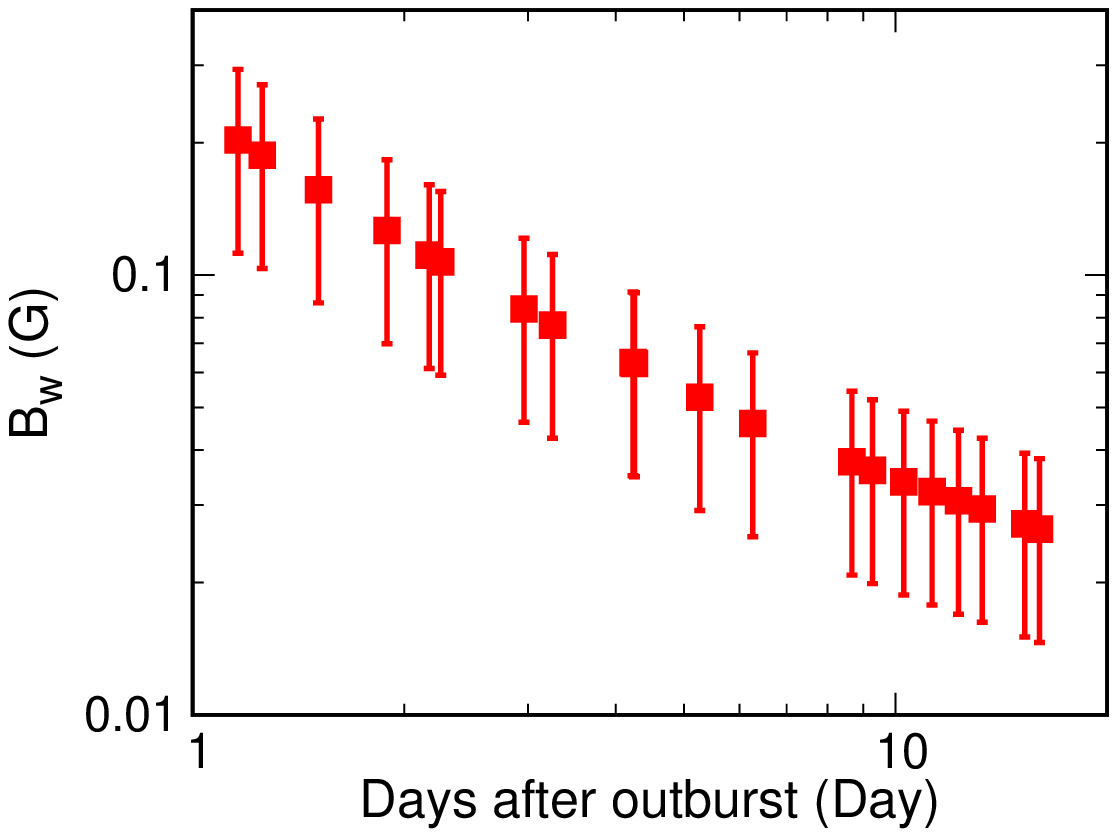}\\
\caption{Top: the density of the RG  wind as a function of time. Bottom: the magnetic field of the RG wind as a function of time.}
\label{fig:rho_B_w_models}
\end{figure}

From $\dot{\text{M}}_{\text{RG}}/$u$_{\text{RG}}$, we  obtain $\rho_{w}$ using Equation \ref{eq:rho}. Figure \ref{fig:rho_B_w_models}  shows the temporal evolution of $\rho_{w}$, where $\dot{\text{M}}_{\text{RG}}/$u$_{\text{RG}}$=$(5\pm 1)\times$10$^{13}$ g cm$^{-1}$. Comparing Figure  \ref{fig:rho_B_w_models} and Figure \ref{fig:Vol_ne_rho_M_models}, we see that $\rho_{s}$ is not 4 times $\rho_{w}$. As mentioned in section 5, we overestimate the volume of the shocked plasma and, for this reason, $\rho_{s}$ is a lower limit. If the average density were representative of the local density and $\rho_{s}$ were 4 times $\rho_{w}$, the volume of the shell between the reverse and the forward shock would be between $\left[ 200-40\right] $ times smaller than the Vol$_{s}$.

As explained in \citet{Tatischeff2007} for RS Oph, a relatively large magnetic field of stellar origin is expected to preexist in the RG wind. Assuming that turbulent motions in
the wind amplify the magnetic field B up to the equipartition value \citep{Bode1985}, the magnetic pressure (P$_{m}$=B$_{w}^{2}/8\pi$) is equal to the thermal pressure (P$_{th}=\rho_{w} k $T$_{w}/ \mu m_{H}$), 
and then B$_{w}$ is given by
\begin{equation}\label{B_w}
B_{w}=\left(8 \pi \rho_{w} k T_{w}/ \mu m_{H} \right)^{0.5}
\end{equation}
where T$_{w}$=10$^{4}$ K is the wind temperature, assumed to be uniform in the RG wind. In the vicinity of the shock front, further magnetic field amplification due to interaction between accelerated particles
and plasma waves is expected to occur \citep{Lucek2000}. Therefore, assuming a time independent amplification factor  $\alpha_{B}$, the magnetic field ahead of the shock will be $B_{0}=\alpha_{B}B_{w}$. 

Figure \ref{fig:rho_B_w_models} (bottom panel) shows the RG wind magnetic field temporal evolution, obtained from equation \ref{B_w} using the derived $\rho_{w}(t)$ for V745 Sco (see Figure \ref{fig:rho_B_w_models}, 
top panel). The magnetic field at the shock of V745 Sco obtained by \citet{Kantharia2015} was B$_{0}$=0.03 G. This value is similar to the B$_{w}$ obtained in this study, and therefore the magnetic field  amplification 
$\alpha_{B}$ should have been small.

\section{Comparison with RS Oph }

RS Oph and V745 Sco belong to the group of long-orbital period novae, which also includes T CrB and V3890 Sgr \citep{Anupama2008}. They are symbiotic systems composed of a massive WD, close to the Chandrasekhar mass ($\sim$1.4 M$_{\odot}$), orbiting a RG companion. The RGs of their systems are classified as M under the Harvard spectral classification \citep[][and references therein]{Russell1914}.   Their distances are very different and widely discussed in the literature. RS Oph seems to be at 1.6 kpc, but there are different studies that place it closer (0.5 kpc \citep{Monnier2006}) or farther away (4.2 kpc \citep{Schaefer2010}) from us.

The orbital period of RS Oph is accepted to be 455.7$\pm$0.8 days. The separation between the WD and the RG is $\sim 1.5$ AU ($\sim 2.4\times10^{13}$ cm) \citep{Fekel2000}. In the case of V745 Sco, the value of the orbital period is not clear.  RS Oph has an irregular recurrence period being its outbursts separated by 9 to 26 years. Taking into account that the last outburst was in 2006 and the shortest recurrent period is 9 years, we can state that at any moment from now an outburst of RS Oph can occur.  On the other hand, V745 Sco has a recurrence period of $\sim$25 years, which was confirmed in the last outburst in 2014.

RS Oph and V745 Sco were observed at practically all the wavelengths in the last outburst. There are two main differences between them. First, V745 Sco was marginally detected by \textit{Fermi}/LAT, while  RS Oph could not be detected because \textit{Fermi} had not been launched yet. Second, there is a larger number of observations of RS Oph than of  V745 Sco,  because the later is fainter and faster. 

We will describe the properties of the plasma of these two novae and their high energy emission, showing possible common features for all the symbiotic systems. More details about RS Oph will appear elsewhere (Delgado \& Hernanz in prep).

\subsection{Plasma properties}

\begin{figure}
		\centering
		\includegraphics[width=0.5\textwidth]{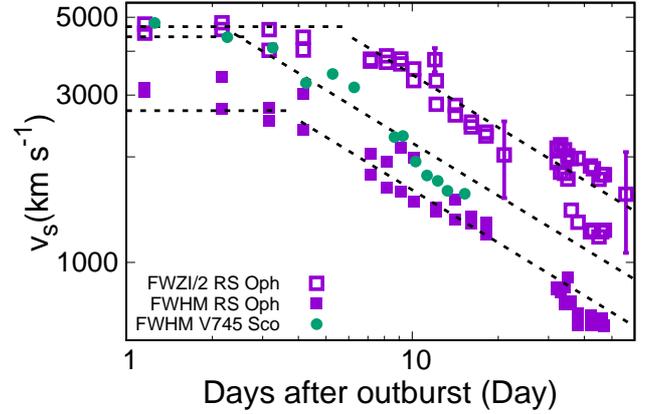}
		\caption{ Temporal evolution of the broad gaussian line component FWHM (solid dots and squares) and FWZI (empty squares) for RS Oph (purple squares  \citep{Das2006, Evans2007}) and V745 Sco (green dots,  same data as Figure  \ref{fig:IRV745Sco_pendientes}).}
		\label{fig:IR_comparado_V745Sco_RSOph}
		\end{figure}

The FWHM and FWZI of spectral lines are  good indicators of the shock velocity. For the spectral lines of  RS Oph, we know their FWHM and FWZI, but for those of V745 Sco, we only know the FWHM.  Figure \ref{fig:IR_comparado_V745Sco_RSOph}	shows the fit of v$_{s}$ derived from the FWZI/2 and the FWHM of the O I and Pa$\beta$ lines of RS Oph \citep{Das2006,Banerjee2009} and the FWHM of the Pa$\beta$ line of V745 Sco. In RS Oph we see that before day 6 v$_{s}$ is practically constant, but after day 6, a clear decrease in the velocity is observed. The power law index obtained is closer to 0.5, associated to the radiative snow-plow phase in SNR, than to 0.3, corresponding to the adiabatic phase (See $\S$\ref{sec:IRobservations}).  In V745 Sco, it is difficult to determine the phases  involved and the day when the slope changes, because uncertainties are unknown. In both novae Phase II was either very short or it did not exist.

\begin{figure}
\includegraphics[width=0.45\textwidth]{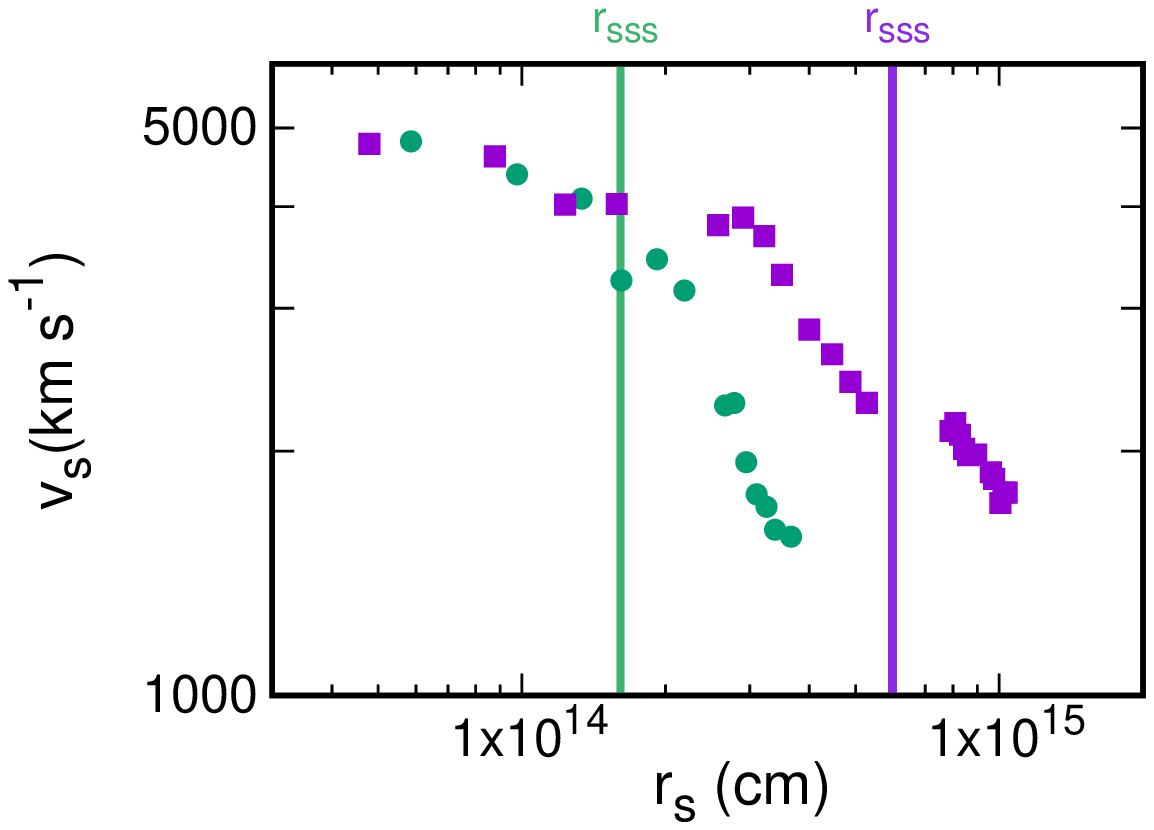}
\includegraphics[width=0.45\textwidth]{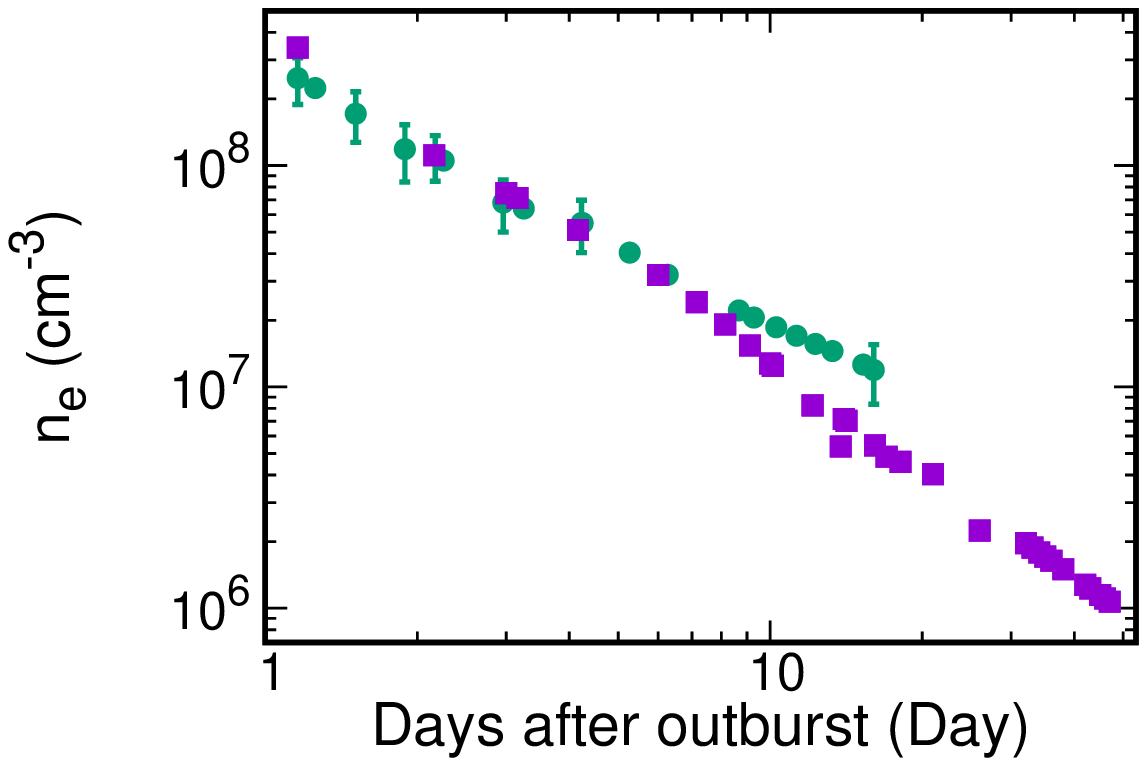}
\includegraphics[width=0.45\textwidth]{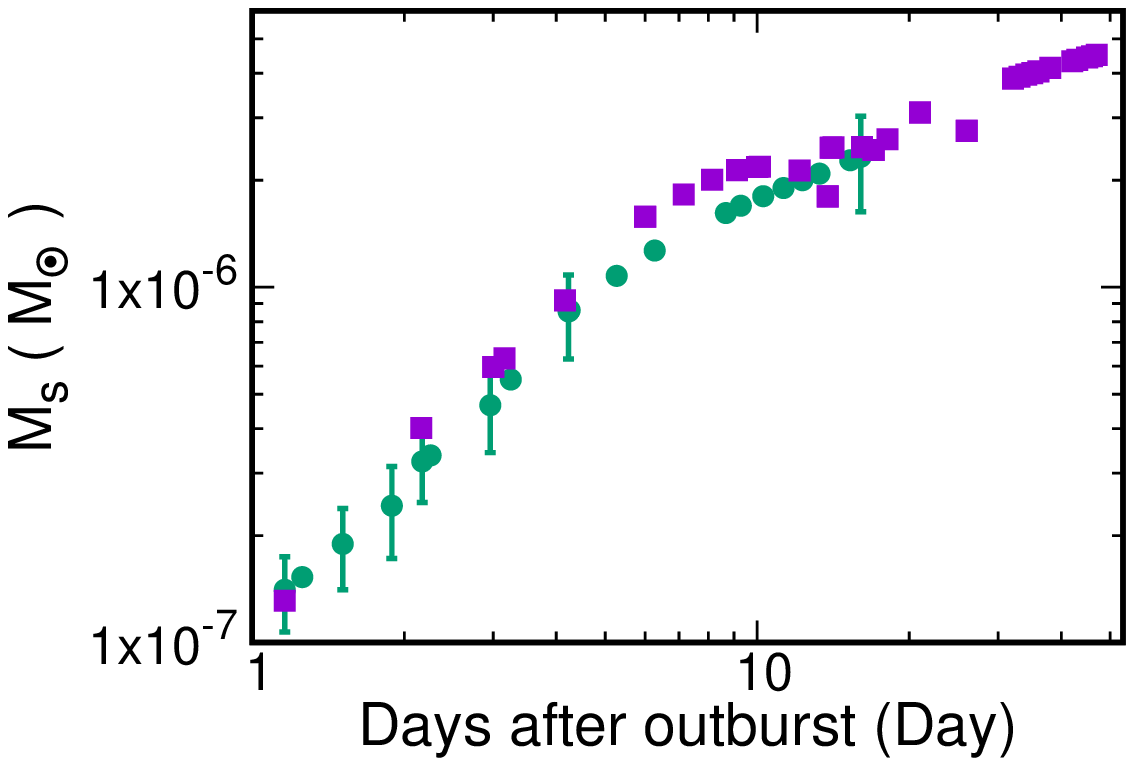}
\includegraphics[width=0.45\textwidth]{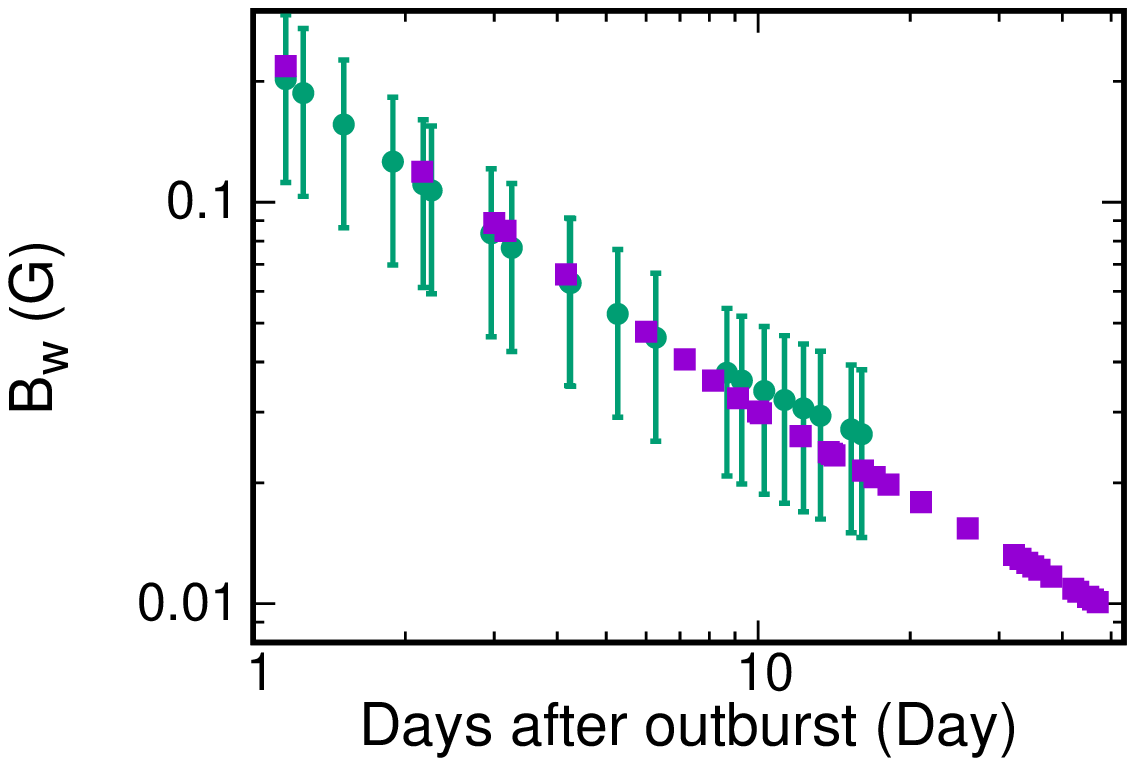}
\caption{ From top to bottom, evolution of the forward shock velocity as a function of the forward shock radius and evolution of the average electron density and the mass of the plasma behind the forward shock; and the magnetic field of the RG wind as a function of time. In all panels purple squares correspond to RS Oph and green dots to V745 Sco. }
\label{fig:R_eje_together}
\end{figure}	

We analyse the temporal evolution of the plasma behind the forward shock to understand better its properties.   We show the forward shock velocity as a function of the forward shock radius in Figure \ref{fig:R_eje_together}.  For V745 Sco,  the large change in the v$_{s}$ slope coincided with the beginning of the SSS phase, when r$_{\text{SSS}}\sim$1.5$\times$10$^{14}$ cm. The large change in the v$_{s}$ slope of RS Oph was on day 6 after the outburst when r$_{s}\sim 4\times10^{14}$ cm. In RS Oph, the location of the forward shock when the SSS phase begins  is equal to r$_{\text{out}}$  (r$_{\text{SSS}}$=r$_{\text{out}}\sim$6$\times10^{14}$ cm, 26 days after the outburst). The top panel in Figure \ref{fig:R_eje_together} shows that there is a small second plateau between 5 and 8$\times10^{14}$ cm, around day 30 after the outburst. In the case of V745 Sco, until day 16 after outburst, the forward shock radius is smaller than r$_{\text{out}}$ ($\sim 8\times10^{14}$ cm). We do not have IR information after day 16. Thus, we do not know if V745 Sco has a plateau close to r$_{\text{out}}$.

As we can see in Figure \ref{EM_flux},  the EM of V745 Sco was practically constant all the time, whereas the volume increases with time.  In RS Oph, the EM was practically constant before day 6 and it decreased with time later on, so n$_{e}$ decreased faster after day 6 (Delgado \& Hernanz in prep.). This implies that after day 6, n$_{e}$ of RS Oph was lower than n$_{e}$ of V745 Sco (Figure \ref{fig:R_eje_together}). 

The mass behind the forward shock can be defined as M$_{s}$=$\rho_{s}\times$Vol$_{s}$, for a hydrogen plasma. Therefore, the differences in the velocities (which propagate to the volume) and in the EMs (which propagate to the $\rho_{s}$)  compensate each other, making the M$_{s}$ in both novae practically equal (Figure \ref{fig:R_eje_together}). M$_{s}$ obtained 1 day after the outburst was the same in both cases ($\sim 1 \times$10$^{-7}$ M$_{\odot}$).

V745 Sco and RS Oph are very similar, but the beginning of the SSS phase is different. Why this phase started so early in V745 Sco is something that is not clear. We expected some clear differences in the masses behind the forward shock, but as we see in Figure \ref{fig:R_eje_together}, RS Oph and V745 Sco look very similar in practically all the properties. However, they have very different ejected masses. RS Oph ejected mass (10$^{-6}$ M$_{\odot}$ \citep{Das2006}) is larger than the ejected mass of V745 Sco (10$^{-7}$ M$_{\odot}$ \citep{Page2015}). Because they are in a symbiotic system and there is the RG wind around the system, the mass behind the forward shock obtained for the first day is not a good diagnosis of the ejected mass. Otherwise their ejected masses would be the same (M$_{s}$=10$^{-7}$ M$_{\odot}$ Figure \ref{fig:R_eje_together}). The influence of the RG wind material is more important than expected the first days after the outburst and it would be interesting to study  its characteristics  more in depth. It can also be related to the geometry of the system, which in  V745 Sco generates an irregular orbital period and  can also be related to the fact that v$_{s}$ of V745 Sco decreased faster than v$_{s}$ of RS Oph.  More information about the system is needed to know the reason of this difference between them.  

In addition to knowing the properties of the plasma through the X-ray observations, we can also get a better understanding of the properties of the RG wind, such as density and magnetic field. \citet{Tatischeff2007}  studied the RG wind density of RS Oph and its mass-loss rate ($(\dot{\text{M}}/ \text{u})_{\text{RG RS Oph}}$=$4\times$10$^{13}$ g cm$^{-1}$). For V745 Sco, we obtain a similar mass loss rate  ($(\dot{\text{M}}/ \text{u})_{\text{RG V745 Sco}}$=$(5\pm 1)\times$10$^{13}$ g cm$^{-1}$).

Regarding B$_w$, it follows the same behaviour in V745 Sco as in RS Oph. However, comparing the amplification produced by the interactions between accelerated particles and the shocked plasma, we get that the amplification of RS Oph ($\alpha_{B}\sim$2 \citep{Tatischeff2007}) was larger than in V745 Sco ($\alpha_{B}\sim$1).  A possible explanation for this difference is that in V745 Sco there was less ejected mass to produce the shock and therefore there were less interactions and less particles accelerated. \citet{MartinDubus2013} state that the low densities in novae  are related with a short $\gamma$-ray emission.  This fact can explain the short duration of the $\gamma$-ray emission in V745 Sco ($\sim$ 1 day),  and the similarities between B$_{w}$ and B$_{0}$  ($\alpha_{B}\sim1$).

\subsection{High energy $\gamma$-rays}

\begin{figure}

\includegraphics[width=0.5\textwidth]{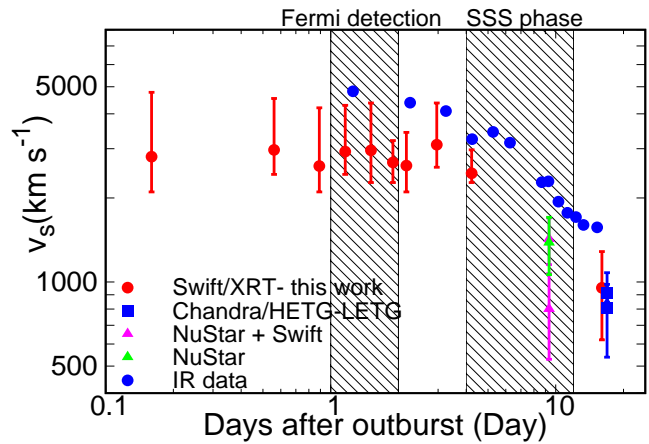}
\caption{Temporal evolution of the shock velocity in IR (see Figure \ref{fig:IRV745Sco_pendientes}) and X-rays (data obtained from Figure \ref{Temperature_V745Sco}).  Shaded areas show the \textit{Fermi} detection and the SSS phase duration. \textit{Swift} measurements have 1$\sigma$ error bars (At 3 $\sigma$, only lower limits are obtained). \textit{NuStar} and \textit{Chandra} values have 3$\sigma$ error bars.}
			\label{fig:IRV745}
\end{figure}

A comparison between the shock velocity estimated from the relation for a test-particle strong shock,  $kT_{s}=\dfrac{3}{16}\mu m_{H} v_{s}^{2}$, and from the width of IR emission lines is shown in Figure \ref{fig:IRV745} for V745 Sco. Shock velocities derived from X-ray data are slightly smaller than those derived from IR, but there are large uncertainties in the X-ray data which prevent us to be conclusive. 

In RS Oph, errors were much smaller and this kind of comparison was used, for the first time, to deduce that particle acceleration was at the origin of the differences between both shock velocities (see Figure 1 in \citet{Tatischeff2007}). As mentioned in \citet{Tatischeff2007}, test-particle $v_{s}$-$kT_{s}$ relationship is known to underestimate shock velocities when particle acceleration is efficient, because the post shock temperature - $T_{s}$ derived from X-ray observations - can be much lower than the test-particle value because of the escape of accelerated particles.

For V745 Sco, since there are no large differences between both values of shock velocities (see Figure \ref{fig:IRV745}), it can be deduced that there was no particle acceleration. However, we must keep in mind that in the case of RS Oph FWZI/2$\geq$FWHM (see Figure \ref{fig:IR_comparado_V745Sco_RSOph}). We do not know if in the case of V745 Sco this difference exists, because we only know this value for one day  \citep{Banerjee2014ATel}. Nevertheless
if it were like that, the real discrepancy between v$_s$ from IR and v$_s$ from X-rays would be larger than what is shown in Figure \ref{fig:IRV745}. Additionally, X-ray data have large uncertainties the first hours after the outburst. Only with better data, the method successfully employed for RS Oph by \citet{Tatischeff2007} could be used for V745 Sco, and the corresponding high-energy emission be deduced and compared with Fermi observations. This comparison was not possible for RS Oph because Fermi had not been launched yet in 2006. 

\begin{figure}
		\centering
		\includegraphics[width=0.5\textwidth]{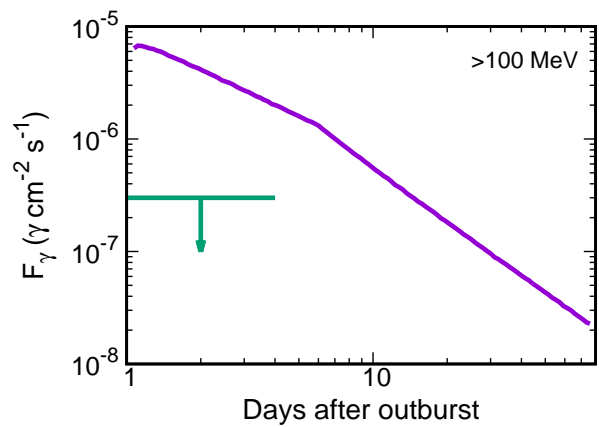}
		\caption{ High energy $\gamma$-ray light curve expected for RS Oph (2006) (purple line) compared with marginal detection of  Fermi/LAT for V745 Sco (green line) \citep[adapted from][]{Hernanz2012}). }
		\label{fig:Fluxgamma_comparado_V745Sco_RSOph}
		\end{figure}

Fermi/LAT detected marginally (2$\sigma$ -3 $\sigma$) V745 Sco; there was no spectrum published, so that the nature of the emission (hadronic or leptonic) could not be determined.  On the other hand, for RS Oph, theoretical spectra and light curves were obtained for E$>$100 MeV, showing that $\pi^0$  decay was dominant over Inverse Compton \citep{Tatischeff2007}.  Figure \ref{fig:Fluxgamma_comparado_V745Sco_RSOph} shows the high energy $\gamma$-ray light curve predicted for RS Oph compared with the flux detected with \textit{Fermi} for V745 Sco.  The factor between their $\gamma$ fluxes is $\sim$ 25 and the distances to V745 Sco and RS Oph are $\sim$8 kpc and $\sim$1.6 kpc respectively. Therefore, the difference in the fluxes can be explained by the differences in their distances.

If there was particle acceleration, both mechanisms, $\pi^{0}$ decay and Inverse Compton, were involved in the $\gamma$-ray emission.   Taking into account the study by \citet{Hernanz2012} and that the ratios $\dot{\text{M}}_{\text{RG}}/$u$_{\text{RG}}$ of RS Oph and V745 Sco are similar, the hadronic process is more likely to be dominant. In the future, one could expect to get the $\gamma$-ray spectrum of V745 Sco and to know the nature of its high energy $\gamma$-ray emission using our study.

\section{Conclusions}

Motivated by the marginal detection of the symbiotic recurrent nova V745 Sco by \textit{Fermi}/LAT, we carried out a multiwavelength analysis of V745 Sco the first days after its outburst in 2014. We have focused on the reduction and analysis of the X-ray observations obtained by \textit{Swift}/XRT. Thanks to the fast repointing  capability of \textit{Swift}, it has been possible  to study the evolution of the properties of the hot plasma the first two weeks after the outburst, in order to progress on the understanding of particle acceleration in the shock.  The fast evolution of the nova did not allow to observe non-thermal X-ray emission with NuStar. As for RS Oph, the differences between the shock velocities deduced from X-rays and IR observations could only be understood if a fraction of the shock energy went into cosmic rays \citep{Decourchelle2000}. Trying to relate the evolution of the hot plasma with the very high energy emission of V745 Sco, we deduce that, again as for RS Oph \citep{Tatischeff2007},  the fact that V745 Sco skipped the adiabatic phase going directly to the radiative phase, most probably indicates the acceleration of particles. From the absorption produced by the RG wind and the expelled material therein, we deduce that the magnetic field amplification is very small and that the dominant process in the particle acceleration is most likely the hadronic one. The comparison between RS Oph and V745 Sco shows that they  are very similar. In the future, with better data and applying the same methods presented in this paper, it will be possible to study the nature of the accelerated particles.

Our main results are summarized below:
\begin{enumerate}
\item We have studied all the early evolutionary phases of V745 Sco by means of a detailed analysis of the available observations with different X-ray satellites, as well as the IR data. 

\item  The temporal evolution of the temperature, N$_{\text{H}}$, EM and flux has been  obtained through  the analysis of the X-ray data.  The temperature decreased faster than in an adiabatic shock due to radiative losses. N$_{\text{H}}$ reached the value of $[N_{\text{H}}]_{\text{ISM}}$ earlier than expected because, probably the SSS emission photoionized the wind material, reducing significantly its opacity.  

\item We have derived the temporal evolution of R$_s$, n$_e$ and M$_s$, with v$_s$ obtained from IR data and EM from X-ray data. 

\item Through the study of the absorption produced by the RG wind ([N$_{\text{H}}]_{w}$), we obtained V745 Sco $\dot{\text{M}}_{\text{RG}}$ which is similar to that of RS Oph.   The derived B$_{w}$ for V745 Sco  is only slightly larger than that obtained in radio, implying a small magnetic field amplification due to particle acceleration.

\item We deduce that the forward shock evolution of V745 Sco skipped the adiabatic phase, as in the case of RS Oph. The fast cooling is related with particle acceleration, which in turn originates the $\gamma$-ray emission detected by Fermi. We obtain that the difference between the F$_{\gamma}$ from V745 Sco and RS Oph can be explained with the differences in their distances. 

\item We could not prove the nature of  particle acceleration in V745 Sco due to the low quality or even lack of data. 

\end{enumerate}

During the next eruption of V745 Sco ($\sim$2039?), it would be interesting to have early observations with a new generation of instruments in the MeV-GeV range such as \textit{e-ASTROGAM} or \textit{AMEGO} \citep{Astrogam2017}.  This instrument would be able to disentangle between the $\pi^{0}$ decay and the Inverse Compton mechanisms. In addition, regarding the radio emission, it would be interesting to observe V745 Sco with the next generation \textit{VLA} and with \textit{SKA} \citep{Brien2015} to obtain information about the thermal and the non-thermal emission the first days after the outburst and also to have enough spatial resolution to resolve the structure of the ejecta.

\section*{Acknowledgements}

This research was supported by the Spanish MINECO grant ESP2017-82674-R and  FEDER funds. LD also thanks the doctoral fund from MINECO BES-2012-056640.


\end{document}